\documentclass{article}
\usepackage[backend=biber]{biblatex}
\usepackage{authblk}
\title{Extreme mid-infrared field enhancement and anapoles in high-index plasmonic metamaterials}
\author[1]{Zoltan Sztranyovszky}
\author[2]{Nicolas Spiesshofer}
\author[2]{Caleb Todd}
\author[2]{Rakesh Arul}
\author[2]{Yeeun Roh}
\author[1]{Rohit Chikkaraddy}
\author[2]{Jeremy J. Baumberg}
\author[1]{Angela Demetriadou}
\affil[1]{School of Physics and Astronomy, University of Birmingham, Birmingham, B15 2TT, United Kingdom}
\affil[2]{NanoPhotonics Centre, Cavendish Laboratory, University of Cambridge, Cambridge, CB3 0US, United Kingdom}

\usepackage{graphicx}
\usepackage{float}
\usepackage[margin=2cm]{geometry}
\usepackage{siunitx}
\usepackage{amsmath}
\usepackage{physics}
\usepackage{hyperref}
\hypersetup{
    colorlinks=true,
    linkcolor=blue,
    filecolor=magenta,      
    urlcolor=cyan,
}
\usepackage[figure]{hypcap}
\usepackage{caption}

\newcommand{\EF}{\mathrm{EF}}

\begin{document}

\maketitle

\begin{refsection}

\begin{abstract}
    High-refractive-index materials underpin a wide range of optical technologies, including communications, imaging, lasers, and integrated photonic systems. 
    Here, we demonstrate a self-assembled metamaterial platform based on gold nanoparticle aggregates with nanometer-scale gaps exhibit remarkably high effective refractive indices exceeding 15 in the mid-infrared regime, while simultaneously producing gap-field enhancements of at least two-orders of magnitude.
    This combination of high refractive index and extreme field enhancement enables exceptionally strong light-matter interactions. 
    We demonstrate this by designing a compact high-index metamaterial device supporting an anapole, which further enhances the nanogap field. 
    By placing quantum emitters with terahertz transitions inside the plasmonic gaps, we show a stimulated-emission response enhanced by at least three orders of magnitude, highlighting applications in non-linear optics, frequency up-conversion and vibrational strong coupling. 
\end{abstract}


\section{Introduction}
\label{s:intro}
Metamaterials (MMs) enable the engineering of emergent optical properties that are not inherent to their individual building blocks, or meta-atoms. A wide variety of MM devices with a myriad of unique electromagnetic responses have been demonstrated, including artificial magnetism, negative refraction and electromagnetic cloaking~\cite{veselago_electrodynamics_1967,pendry_magnetism_1999, smith_negative_2000,pendry_negative_2000,pendry_controlling_2006,kubo_tunability_2007}. 
These effective properties arise from the homogenized response of subwavelength meta-atoms and are governed primarily by their geometry and lattice arrangement~\cite{mallet_maxwell-garnett_2005,aspnes_plasmonics_2011}.
MMs were first widely exploited in microwave antennas systems, where millimeter-scale fabrication is readily accessible.
In this regime, they have transformed antenna functionality by enabling miniaturization, gain enhancement and beam shaping~\cite{milias_metamaterial-inspired_2021,tadesse_application_2020}. 
Extending these concepts to the mid-infrared (MIR) and visible regimes is highly desirable for applications in biosensing, photocatalysis, lasers, integrated optics and signal processing~\cite{cui_roadmap_2024,kuznetsov_roadmap_2024,schulz_roadmap_2024}.
Such devices can benefit from MMs with effective high-refractive indices and low losses, but this is increasingly difficult to achieve at higher frequencies due to fabrication and material limitations, placing increasingly stringent demands on them~\cite{yoon_challenges_2016,leng_meta-device_2024}.

Current approaches to MIR and visible MMs rely predominantly on top-down fabrication methods, such as electron-beam lithography and direct laser writing~\cite{chen_nanofabrication_2015,harinarayana_two-photon_2021}.
These methods offer excellent control over meta-atom geometry and spatial arrangement, enabling intricate architectures with tailored optical responses, including effective chirality and non-reciprocity, and have been used in applications for sensing, nonlinear enhancement and optical beam control~\cite{wang_optical_2016,asadchy_bianisotropic_2018,yang_advanced_2024,tsarapkin_double_2025,li_magneto-optical_2024}. 
However, their reliance on serial or highly specialized fabrication processes often makes them costly, time-consuming and difficult to scale. 
In addition, many MM responses tend to be inherently narrow-band, dispersive and lossy, since the desired optical properties typically arise near resonances, which imposes further constraints on device performance. 
Dynamically retuning these resonances after fabrication remains challenging, although active and mechanically reconfigurable assembly offers an a promising pathway~\cite{zhao_mechanically_2024,jung_rise_2024}.
Bottom-up self-assembly provides an attractive alternative, in which chemical interactions drive the formation of metamaterial structures, and is comparatively inexpensive, rapid, scalable and easy to implement~\cite{grzelczak_directed_2010,flavell_programmed_2023}.
In this context, nanoparticle (NP) aggregates are particularly promising: they can exhibit broadband high effective refractive indices~\cite{chung_optical_2016,doyle_tunable_2018,kim_metal_2018,palmer_extraordinarily_2019,huh_soft_2020,kim_achieving_2024} while remaining compatible with self-assembly routes. 
Nevertheless, previous studies have focused mainly on their bulk optical response. 
Recent advances in self-assembly now allow plasmonic gaps to be controlled down to the nanometer scale~\cite{arul_giant_2022,spiesshofer_tailoring_2025,baumberg_extreme_2019,zhu_quantum_2016}, creating an opportunity to combine high-index behavior with extreme local-field enhancement, a feature that remains largely unexplored.

Here we show that multi-layer aggregates (MLaggs), made of gold NP and structured with nanometer-sized gaps, simultaneously support ultra high effective refractive indices exceeding 15  over a broad MIR regime, and extreme local field enhancement of at least 2-orders of magnitude. 
We find that the high refractive index remains nearly dispersionless throughout the MIR, where fields can couple directly to molecular vibrational states. 
The effective refractive index can be tuned by varying the NP size and/or morphology, the gap size, aggregation and gap refractive index.
We exploit these properties to design a compact high-index MM resonator that supports an anapole~\cite{miroshnichenko_nonradiating_2015,papasimakis_electromagnetic_2016}, a non-radiative state associated with strong field confinement inside the resonator~\cite{totero_gongora_anapole_2017,zhang_anapole_2020}. 
By combining the confinement provided by the nanogaps with the confined fields of the anapole, we obtain exceptionally large field enhancements that can be tuned across the MIR regime. 
Finally, by placing quantum emitters inside the nanogaps, we demonstrate light amplification with the MM resonator, establishing self-assembled NP aggregates as a high-index, scalable platform for extremely enhanced light-matter interactions. 

\section{Results and Discussion}
\label{s:ema}

\subsection{Refractive index, loss, and field enhancement}
\label{ss:rad_and_gap}
\begin{figure}
    \centering
    \includegraphics[width=0.9\linewidth]{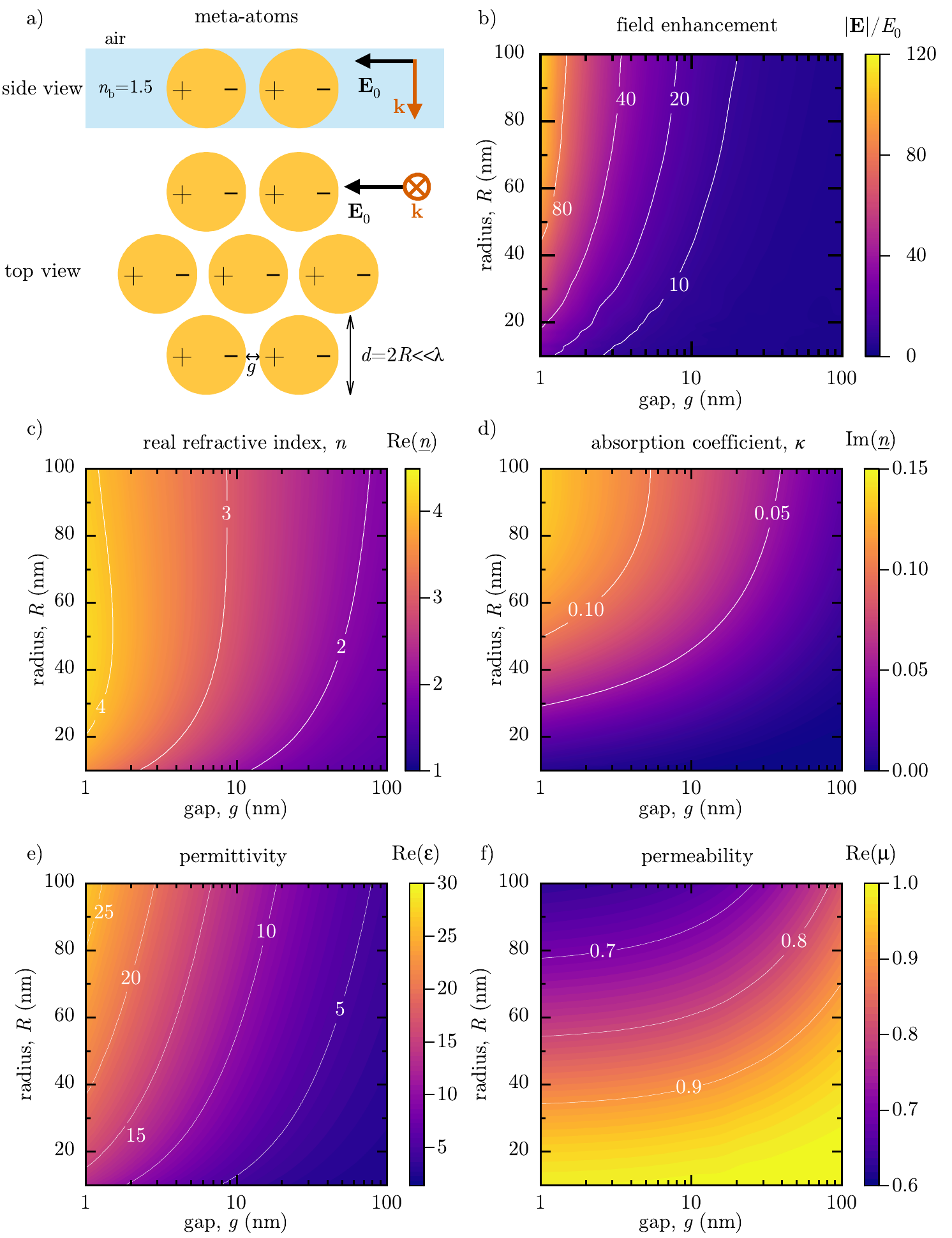}
    \caption{a) Schematic illustration of the spherical NP aggregate MM, embedded in a medium of $n_b=1.5$. For various NP radii and gap sizes, we show the: b) field enhancement on the surface of the gold NPs; 
    c) the real and d) imaginary effective refractive index $n$ and $\kappa$; e) the real electric permittivity $\Re(\varepsilon)$; f) real magnetic permeability $\Re(\mu)$. All values are taken at $\lambda=10\unit{\um}$, where they are nearly dispersionless. }
    \label{fig:n_k_E}
\end{figure}

Previous works have shown that NP aggregates produce an effective refractive index controlled by the gap size and morphology~\cite{chung_optical_2016,doyle_tunable_2018,kim_metal_2018,palmer_extraordinarily_2019,huh_soft_2020,kim_achieving_2024,spiesshofer_tailoring_2025}.
Using this knowledge, we consider a MM made of a single layer of gold spherical NPs of radius $R$, separated by gap $g$, arranged in a hexagonal lattice forming a slab of thickness $d=2R$, and embedded in a medium of $n_\textrm{b}=1.5$, see~\autoref{fig:n_k_E}.a. 
The MIR effective refractive index and field enhancement in the gaps are primarily determined by the radius of the NPs ($R$) and the gap size ($g$).
The size of the NPs is usually set during their synthesis~\cite{ruan_growth_2014,zheng_successive_2014}, while the gap size that can reach sub-nm scales, is controlled by the molecules binding the NPs together~\cite{doyle_tunable_2018,arul_giant_2022}, which also define the refractive index of the host medium.
For example, NP aggregates of radius $R=50\unit{\nm}$ and gap of $g=1\unit{\nm}$, lead to a MM with an effective refractive index of $n\approx4$ throughout the MIR regime $\lambda > 1.5\unit{\um}$~\cite{spiesshofer_tailoring_2025}.
. The same MM leads to field enhancement in the nanogap of the order of $EF=|\vb E|/E_0 \approx 90$.
Since the MM effective optical properties are mostly dispersionless throughout the MIR regime, we show in~\autoref{fig:n_k_E}.b-d the complex effective refractive index and field enhancement at $\lambda=10\unit{\um}$, which represents the overall MIR behavior of the MM for various NP radii and gap sizes. 
The field enhancement inside the nanogaps is stronger for the smaller gaps, due to stronger field confinement, and larger NPs, since larger NPs provide increased polarizability, as shown in~ \autoref{fig:n_k_E}.b. 
At the same time, the real effective refractive index $n$ and the electric permittivity $\varepsilon$, shown in~\autoref{fig:n_k_E}.c and e respectively, follow similar increasing trends for smaller gaps and large NPs, which is also consistent with the Maxwell Garnett approximation~\cite{mallet_maxwell-garnett_2005, garnett_colours_1904, markel_introduction_2016} (large NPs and small gaps lead to a high filling factor $f$, and thus an increase in the effective permittivity). 

In a certain regime ($R>50\unit{\nm}$ and $g<2\unit{\nm}$) however, the effective refractive index $n$ essentially plateaus. 
This is due to a diamagnetic effect emerging at the long wavelength limit, since opposing magnetic fields are generated by the induced currents in each NP~\cite{pendry_magnetism_1999,chung_optical_2016,westerberg_is_2025}, which is often overlooked.
It leads to an effective magnetic permeability of $0<\mu<1$ (see \autoref{fig:n_k_E}.f), and can significantly impact the value of the effective real refractive index since $n=\Re(\sqrt{\varepsilon \mu})$.
Also, these induced currents are the source of the effective imaginary refractive index $\kappa$ (i.e. loss), shown in~\autoref{fig:n_k_E}.d, where larger induced currents lead to higher losses (see SM S.II~\cite{sztranyovszky_supplementary_2025}). 
This behaviour ($n$ plateaus and $\kappa$ increases) is better understood if one takes a Taylor expansion for the real and imaginary parts of the refractive index (since $\Re(\varepsilon)\gg \Im(\varepsilon)$ and $\Re(\mu)\gg \Im(\mu)$), which leads to the real part of the effective index $n\approx\sqrt{\Re(\varepsilon)\Re(\mu)}$ and the imaginary part of the  effective index $\kappa\approx \left[\text{Re}(\varepsilon)\text{Im}(\mu) + \text{Im}(\varepsilon)\text{Re}(\mu)\right]/2n$, with the second term dominating the value of $\kappa$ in this particular system (see SM S.III for further details). 
We note that the refractive index of the host medium $n_\textrm{b}$ also impacts the effective refractive index (see SM S.IV), which is also consistent with the Maxwell Garnett approximation~\cite{mallet_maxwell-garnett_2005,garnett_colours_1904, markel_introduction_2016}.
Nevertheless, there is an optimum regime for the MM's desired optical properties, depending on applications. For example, if one requires to have a high refractive index $n$ with low loss $\kappa$, then NPs of $R=20\unit{\nm}$ and gap size of $g=1\unit{\nm}$ lead to effective $\underline{n}\approx 4+i0.025$. 
However, if one also requires high field enhancement in the nanogaps, then larger NPs of $R=50\unit{\nm}$ and gap size of $g=1\unit{\nm}$ lead to  $\underline{n}\approx4.25+i0.1$ and $\EF\approx90$, on which we focus for the rest of the paper. 

Additionally, one can further extend from single-layer to multi-layer NP aggregate structures to create MLaggs. These further increase the effective refractive index by $5-10\%$ for tightly packed layers~\cite{spiesshofer_tailoring_2025}. 
This is also consistent with the Maxwell Garnett approximation, since tightly-packed NP layers lead to higher filling fractions. The increased thickness of the MLagg also leads to Fabry-P\'erot (FP) resonances forming, which shift to longer wavelengths (across the MIR regime) for larger MM slab thicknesses (see SM S.V), allowing simple tuning of the resonance position.

\subsection{Ultra-high refractive index aggregate metamaterials}
\label{ss:shape}

\begin{figure}
    \centering
    \includegraphics[width=0.95\linewidth]{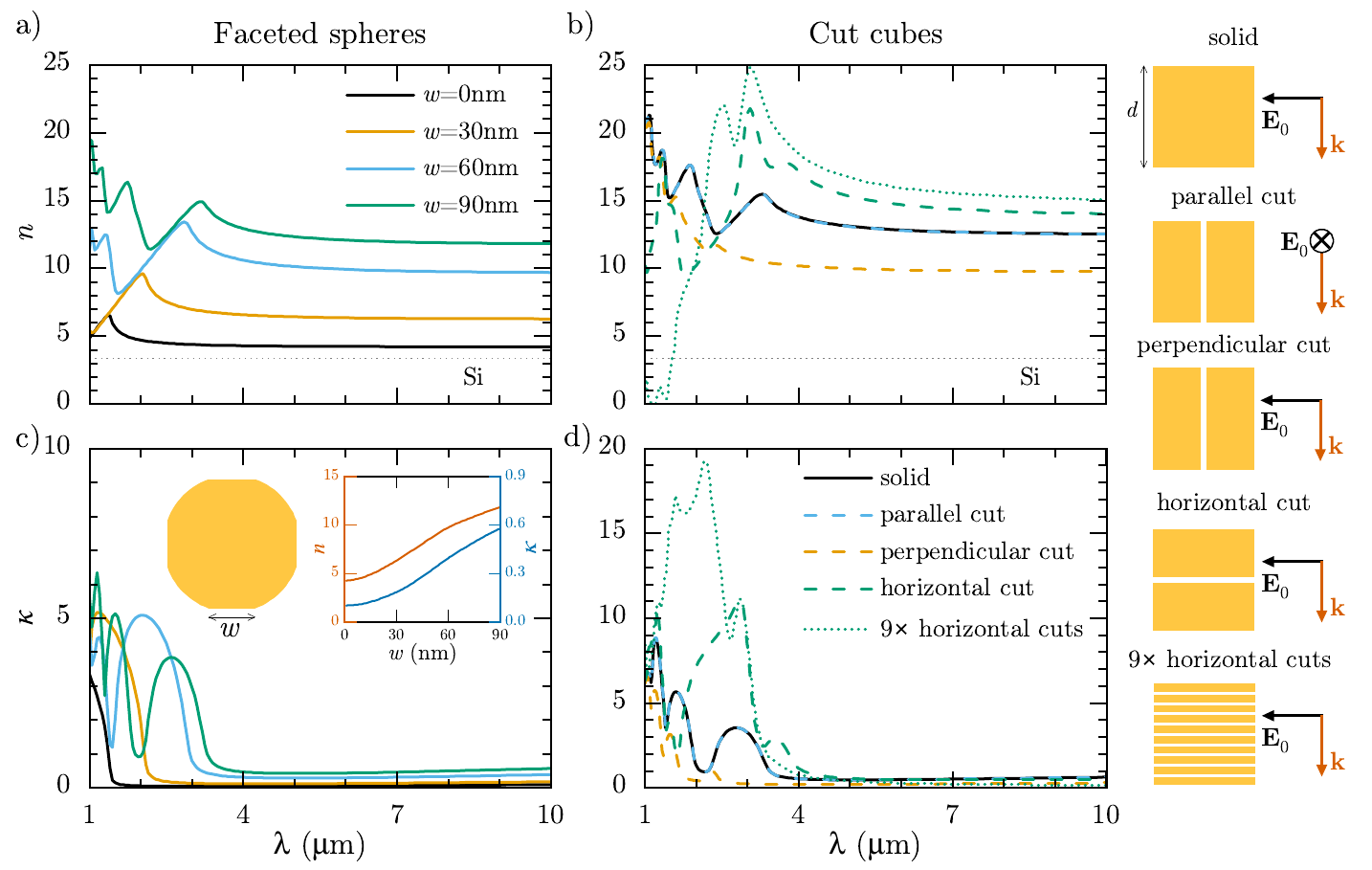}
    \caption{ Effective refractive index $n$ of a) faceted spherical NP aggregates of facet width $w$, and b) cubical NP aggregates of various cuts across the NP. The refractive index of silicon is indicated by a thin gray dotted line for reference. The effective loss $\kappa$ for c) faceted spherical NP aggregates, with  the left inset showing the schematic illustration of the facet, and right inset showing the parameter dependence at $\lambda=10\unit{\um}$, and d) for cubical NP aggregates of various cuts across the NP, which are shown on the right. All the NPs are embedded in a slab of $n_\textrm{b}=1.5$ and height $d=2R=100\unit{nm}$.}
    \label{fig:n_k_shapes}
\end{figure}

The synthesis of gold NPs has advanced significantly in the last few decades to allow for the formation of complex NP morphologies~\cite{liz-marzan_colloidal_2020}. Nevertheless, due to the crystalline nature of gold, the synthesis of spherical NPs often leads to polyhedral particles with facets of varying sizes that often can be controlled~\cite{kim_metal_2018,spiesshofer_tailoring_2025,bedingfield_multi-faceted_2023,elliott_fingerprinting_2022,lu_continuous_2025}. 
Therefore, we now introduce facets of size $w$ on the spherical NPs, which increases the effective refractive index of the aggregate MMs~\cite{spiesshofer_tailoring_2025} (see \autoref{fig:n_k_shapes}.a), in agreement with an increased filling fraction for the Maxwell Garnett theory. 
As for the spherical NP aggregates, the effective $n$ remains relatively dispersionless throughout the IR regime, with the effective loss $\kappa$ (see \autoref{fig:n_k_shapes}.c) also remaining relatively constant and taking low values. 
It should be noted that even though the effective $n$ increases for larger NP facets, the maximum field enhancement actually reduces, since the electrons are distributed across the facet area for each nanogap. 
The average field enhancement on the surface of the NP increases though (see SM S.VI), since there are more electrons per unit volume.
At higher frequencies, we reach the Brillouin zone of the MM, where Bragg scattering opens a band-gap increasing both $n$ and $\kappa$. 
However, in this regime effective homogenization methods are not easily applicable. 
In the case of faceted MLaggs this higher refractive index leads to redshifted Fabry-Perot modes, in line with the effective photonic band reaching the Brillouin zone at lower frequencies for large $n$ (see SM S.VI). 

To further increase the effective $n$, we consider cubical NP aggregates~\cite{huh_soft_2020,kim_achieving_2024,hoang_ultrafast_2015,wang_quantitative_2020}. We place the cubical NPs on a square lattice to maximize the filling factor of the system, keeping their edge length $100\unit{\nm}$ and gap size $g=1\unit{\nm}$. 
To reflect the fact that synthesized cubic nanoparticles typically lack perfectly sharp edges, we model them with rounded corners using a curvature radius of $1\unit{\nm}$ (see SM S.VI for more information on the impact of the curvature). 
The MM made of aggregated cubical NPs has a nearly identical dispersive behavior as the faceted NP aggregates (see \autoref{fig:n_k_shapes}.b and d).
Fundamentally, the effective dielectric behavior of the MM emerges from the induced dipoles across the nanogaps which remains nearly identical between the cubical NP aggregates and the spherical NPs of diameter $2R=100\unit{\nm}$ and facet size $w=90\unit{\nm}$. 
However, the value of the effective $n$ is higher due to the larger filling factor. 
Notably, this is one of the highest effective $n$ reported in the literature~\cite{huh_soft_2020,kim_achieving_2024}.

To further amplify the effective $n$, we consider cubical NPs with `cuts' of $1\unit{\nm}$ to increase the number of gaps per unit volume~\cite{gao_low-loss_2024}. 
Due to the thin gap the volume fraction of gold in the unit cell only decreases negligibly, but the induced dipoles per unit volume increases. 
We initially consider a `parallel cut' design (see \autoref{fig:n_k_shapes}), where the wavevector $\vb k$ and the incident field $\vb{E}_0$ are both parallel to the `cut' plane. 
This does not change the refractive index, as the induced dipole moment per unit volume and induced current loop remains the same, and thus the permittivity and permeability as well (see SM.VI). 
We then introduce a `perpendicular cut', where the cut plane is perpendicular to $\vb E_0$ and parallel to $\vb k$. 
At long wavelengths, this decreases the induced dipole moment and the permittivity by half, while also impeding the current flow along $\vb E_0$, lessening the diamagnetic effects, and overall reducing the effective refractive index by approximately $20\%$.
At short wavelengths, while both the permeability and permittivity changes significantly, they do so in opposite ways, thus the overall effect leaves the refractive index mostly unchanged.

To maintain this low diamagnetic effect of the `perpendicular cut' design, but without impacting the induced dipole moment per unit volume, we now introduce a 'horizontal cut' that is parallel to $\vb E_0$ and normal to $ \vb k$.
This increases the refractive index at long wavelengths, since the effective permittivity is governed by the induced dipole moment that remains unchanged from the solid cubic design. 
However, the diamagnetic effect is much lower, since the current loop is impeded along the $\vb{k}$-direction~\cite{choi_terahertz_2011} (see SM.VI), which drives the magnetic permeability $\mu$ closer to unity. 
Note that for the `horizontal cut' structure, we essentially have two layers stacked, and at short wavelengths this can affect the coupled resonances between them, increasing or decreasing the refractive index. 
To further reduce the diamagnetic effect, we introduce nine horizontal cuts, creating all together ten thin slices of gold stripes, each with a height of $h=9.1\unit{\nm}$, stacked with $1\unit{\nm}$ nanogaps between them, giving the original thickness $d=100\unit{\nm}$.
The effective refractive index for the `$9\times$ horizontal cuts' design increases to $n\approx15$ in the long wavelength regime, with losses reduced by about a factor of $5$ compared to the solid cube.
At higher frequencies, the effective refractive index of this new design peaks to an exceptionally high value of $n\approx24.9$, which is remarkably high for the near-IR regime.
Within this regime the field enhancement remains substantial ($|\vb E|/E_0 > 40$) and can facilitate enhanced light-matter interactions, such as sensing and Raman spectroscopy. 
In summary, we thus find that the highest refractive indices arise from stacking thin square sheets of gold separated by ultra-thin gaps, producing a highly anisotropic MM, similar to hyperbolic materials~\cite{poddubny_hyperbolic_2013}, epsilon-near-zero structures~\cite{hendrickson_coupling_2018,suresh_enhanced_2021}, and even birefringent materials~\cite{zhu_manipulating_2015}. 

\subsection{Emerging anapoles in metamaterial resonators}
\label{s:anapole}
 
\begin{figure}
    \centering
    \includegraphics[width=0.9\linewidth]{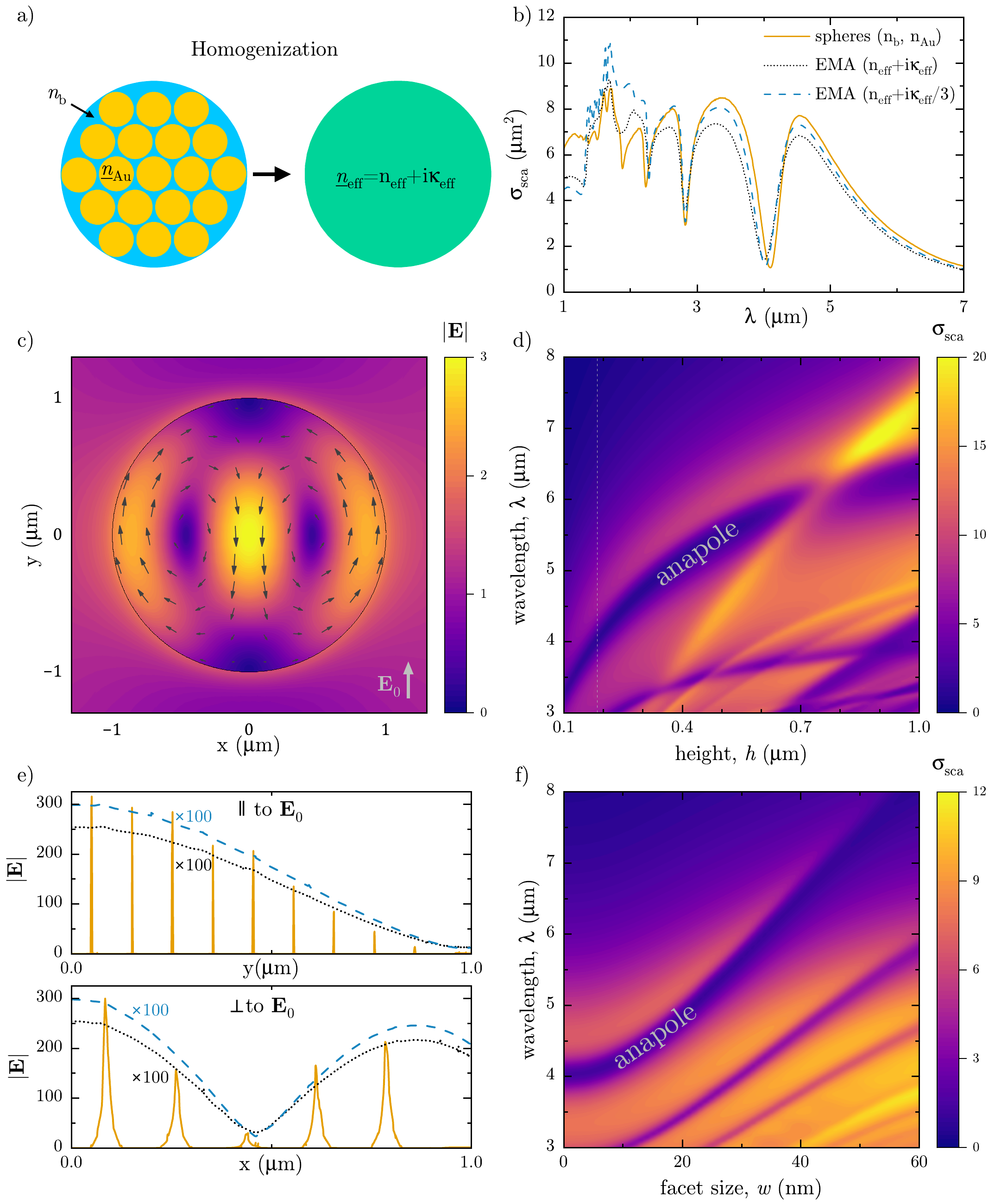}
    \caption{a) Schematic illustration of the MLagg cylindrical disk and its homogenization. b) The scattering cross sections for the inhomogeneous cylindrical resonator made of two layers of gold NP aggregates of $R=50$nm and $g=1$nm (solid gold line), and the homogenized structure with height $h=184$nm, radius $R_\text{cyl}=1$\textmu m, described with the effective refractive index (black dotted line), and the homogenized structure with the loss rate reduced by a factor of 3 (blue dashed line). c) Near field of the homogenized structure at the anapole wavelength $\lambda=4\unit{\um}$. d) Scattering cross section dependence on the height of the cylindrical resonator. Dotted line indicates the height used on b). e) Field enhancement inside the inhomogeneous and the homogeneous cylindrical disk, measured at the center of the first layer of NP aggregates at the anapole wavelength $\lambda=4\unit{\um}$. f) Scattering cross section dependence on the facet size $w$ of the spherical NP aggregates.}
    \label{fig:anapole}
\end{figure}

Using the exceptionally high refractive index and the high field enhancement in the NP gaps, we now use the aggregate MM to build a cylindrical resonator that supports an anapole~\cite{miroshnichenko_nonradiating_2015, baryshnikova_optical_2019,totero_gongora_anapole_2017,yang_anapole-assisted_2018,kuznetsov_special_2022}  in the MIR regime.
Anapoles are non-radiative states, usually observed in high-index dielectric spheres or cylinders, and are characterised by a strong field confinement in the center of the resonator~\cite{papasimakis_electromagnetic_2016,savinov_optical_2019}. 
In Cartesian multipole decomposition they arise explicitly due to the interference of electric and toroidal dipole moments~\cite{babicheva_multipole_2021,miroshnichenko_nonradiating_2015}, where their opposite phase essentially limits radiation to the far-field. 
Anapoles have been successfully used to enhance various non-linear processes~\cite{baryshnikova_optical_2019} and lasing~\cite{totero_gongora_anapole_2017,tripathi_lasing_2021}.
Here, we use the high effective refractive index of the NP aggregate MM to form a cylindrical disk of radius $R_\textrm{disk}=1\unit{\um}$ and height $h=184\unit{\nm}$ (corresponding to two tightly packed layers). 
This MLaggs MM is made of perfect spheres ($w=0\unit{\nm}$) of radius $R=50\unit{\nm}$ and gap sizes $g=1\unit{\nm}$, with the homogenization procedure schematically illustrated in \autoref{fig:anapole}.a. 
This structure allows us to superposition the anapole and nanogap field enhancements, to facilitate unprecedented light-matter interactions in the MIR regime.

We demonstrate the existence of the anapole by obtaining the scattering cross-section spectrum of the cylinder, using the effective refractive index of the MLagg ($n_\textrm{eff}+i\kappa_\textrm{eff}$), shown  in \autoref{fig:anapole}.b (black dotted line). The first anapole emerges at $\lambda=4$\textmu m, where the scattering is minimized, with the field showing the characteristic anapole features in~\autoref{fig:anapole}.c.
The anapole has a strong field enhancement in the center of the cylinder and is very robust with respect to losses~\cite{totero_gongora_anapole_2017,zhang_low_2023}. 
Instead of a homogenized medium, we now form the same cylinder with a two-layer MLagg, and obtain the same spectrum (orange solid line) seen in \autoref{fig:anapole}.b. 
The results are in very good agreement, showing the anapole at the same wavelength.
The differences on the scattering cross-section are mainly due to the finite size of the cylinder and edge effects, which changes the overall effective optical properties of the MLagg compared to an infinite NP aggregate layer~\cite{guerra_effective_2025}. 
The fact that the anapole frequency does not shift significantly indicates that the difference is primarily for $\kappa_\textrm{eff}$. 
We find that reducing loss to $\kappa_\textrm{eff}/3$ mostly recovers the amplitude of both the absorption and scattering cross-sections (blue dashed line) in \autoref{fig:anapole}.b for $\lambda>2\unit{\um}$, where the effective index is non-dispersive (see SM S.VII for absorption cross section).

The field enhancement within the cylindrical MM resonator is shown in \autoref{fig:anapole}.e., for a homogeneous effective medium (black dotted line), homogeneous medium of reduced losses (blue dashed line) and the MLagg (orange full line). 
The overall shape of the field enhancement at the anapole wavelength $\lambda=4 \unit{\um}$  remains similar for the three cases, and follows the anapole field shape. 
The field enhancement for the MLagg anapole though is two orders of magnitude stronger compared to the homogeneous medium anapoles, reaching $EF_\textrm{MLagg}\approx300$ compared to $EF_\textrm{homog}=3$ throughout the cylindrical resonator. 
This extreme field enhancement indicates that there is a superposition of the anapole and nanogap field enhancements, which leads to exceptionally high field confinements, not commonly seen in the MIR regime. 
Notably, the inhomogeneous resonator retains the anapole behavior of the homogenized system with the effective optical properties. Thus homogenization removes the need to model to model the exact near-field profile within the MM, such as in~\cite{yang_theory_2011}, and provides a computationally efficient way to design and optimize MM devices.

While using MLaggs to form an anapole device, one can also tune the anapole wavelength fairly easily. For example, one can change the disk height by simply stacking more MLagg layers, as shown in~\autoref{fig:anapole}.d., or increase the disk radius (see SM S.VII), which shifts the anapole further into the MIR. 
Apart from the cylindrical disk size, one can tune the effective $n$ of the MLagg by controlling the meta-atom geometry and nanogap size.  
For example, increasing the facet width of the spherical NP aggregates leads to increased effective $n$ and a consistent redshift for the anapole to the far-IR regime (see \autoref{fig:anapole}.f).
It should be noted, that although anapole MMs have been proposed before, they tend to be limited to MMs where each meta-atom supports an anapole~\cite{canos_valero_theory_2021,tripathi_lasing_2021,yao_plasmonic_2022,luo_perfect_2023,xing_plasmonic_2024,hassan_anapole_2025}, and not an anapole formed due to the effective properties of the MM. 
This also means that fabrication control requirements are loosened as primarily the micron-scale structure matters for the formation of the anapole. 
Using this effective approach, and combining the boost from the nanogaps and the anapoles, one can reach exceptionally high field enhancements.
This can considerably boost light-matter interactions to significantly enhance vibrational pumping and lasing, and achieve efficient energy up-conversion~\cite{xomalis_detecting_2021,chikkaraddy_single-molecule_2023}.

\subsection{Vibrationally mediated light emission}
\begin{figure}
    \centering
    \includegraphics[width=0.95\linewidth]{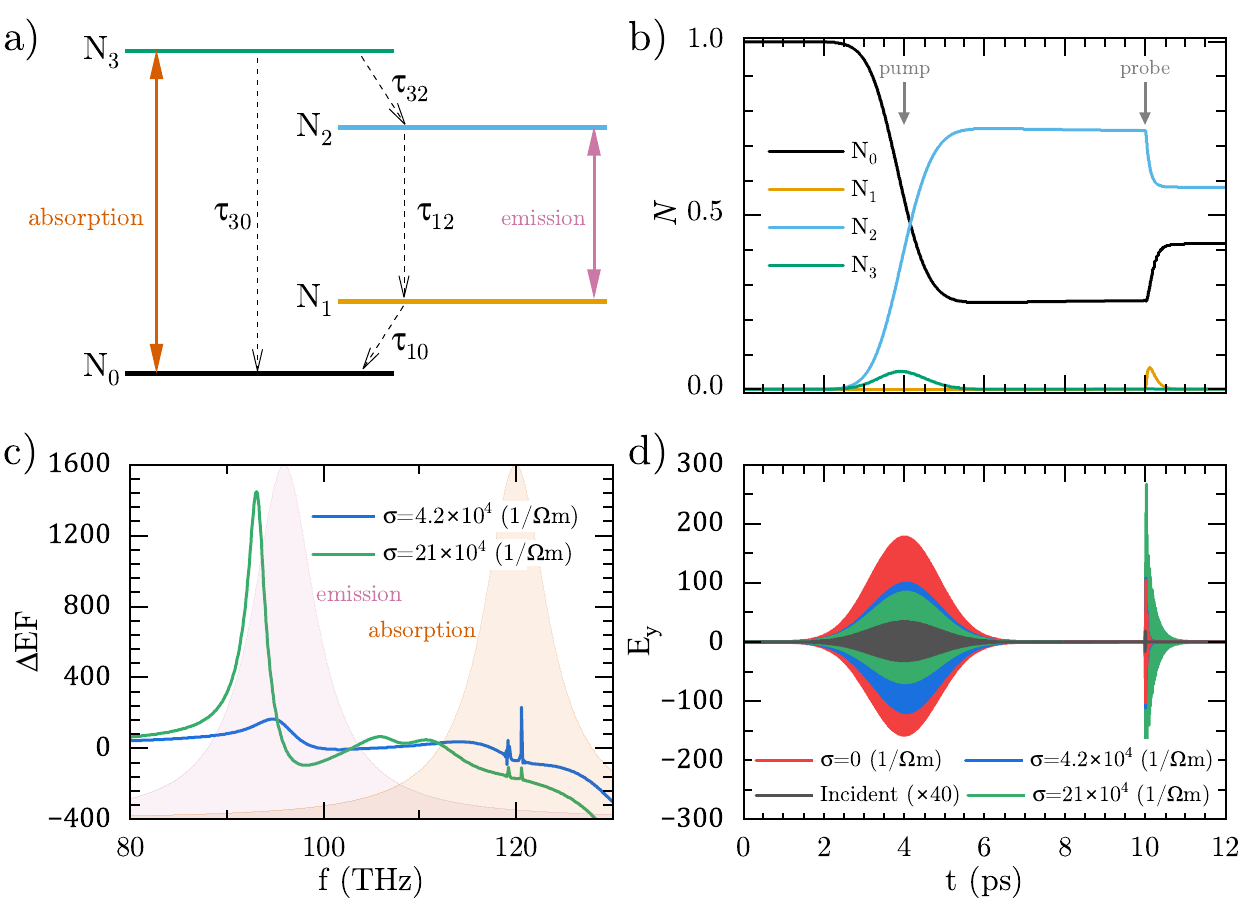}
    \caption{a) Schematic illustration of the four-level system and allowed transitions. Solid arrows represent radiative transitions, and dashed arrows non-radiative transitions. b) Normalized population density of the four level system as a function of time. c) Change of the field enhancement in the frequency domain ($\Delta EF = EF(\sigma\neq0)-EF(\sigma=0)$) due to the presence of the 4-level system. Shaded areas indicates the normalized emission (pink) and absorption (orange) lines of the four-level system. The localized sharp peaks around $f=120\unit{THz}$ are due to numerical noise. d) Time dependent fields along the polarization of the incident wave, inside the central nanogap of the cylindrical resonator, when no materials is present (incident), when only the MM is present ($\sigma=0$), and two different four-level systems with coupling coefficients of $\sigma=4.2\unit{\Omega^{-1}m^{-1}}$ and $\sigma=21\unit{\Omega^{-1}m^{-1}}$.}
    \label{fig:time_domain}
\end{figure}

To take advantage of this exceptional field enhancement, we place quantum emitters in the nanogaps between the NPs, such as quantum dots and semiconducting materials~\cite{hu_robust_2024}, rare-earth lanthanides~\cite{chen_sub-50-ns_2022}, diamond defects~\cite{boyce_plasmonic_2024}, or simply molecules with polar vibrations~\cite{xomalis_detecting_2021}. 
While not all of these materials are suitable to be placed in such small gaps at the moment, the field is progressing extremely fast on this direction~\cite{baumberg_extreme_2019}.
Assuming that we are using polar molecular vibrations, the molecules in the nanogaps experience a combined field enhancement from both the nanogap and anapole field confinement. 
We model the molecules surrounding the NPs of the MLagg as a four-level system described by Maxwell-Bloch equations~\cite{wuestner_gain_2011,valcarcel_semiclassical_2006} (see \autoref{fig:time_domain}.a for schematic illustration, and SM S.VIII for further details), while keeping the host refractive index as $n_\textrm{b}=1.5$. 
We solve this system in the time-domain, and consider only a single layer of NP aggregates to reduce the computational demands of the calculation. 
For this system, the anapole emerges at $\lambda=3125\unit{nm}$, with which we align the emission line of the four-level system. 
The absorption line of the four-level system is aligned with the scattering peak of the system at $\lambda=2500 \unit{nm}$ (see SM S.VIII for spectra). 
The non-radiative decay rates are set to $\tau_\textrm{32}=\tau_{10}=50\unit{fs}$, while absorption and emission decay rates are set to $\tau_\textrm{30}=\tau_{12}=500\unit{ps}$, to allow for population inversion. The dephasing is set to $20\unit{fs}$. We also define the coupling coefficient $\sigma=\mu^2 \rho/\hbar$, where $\mu$ is the dipole moment of the molecule for both the absorption and emission transitions and $\rho$ is the number density per unit volume of the molecules.

We consider a long, narrow frequency pump pulse of $t_\text{pump}=4\unit{ps}$, centered around the absorption line, illuminating and reaching the MLagg disc at $t\approx 4 \unit{ps}$.
This is followed by a short, wide frequency probe pulse of $t_\text{probe}=24\unit{fs}$, centered around the emission line, which reaches the system at $t\approx 10 \unit{ps}$ to characterize its optical response (see \autoref{fig:time_domain}.d black line). 
The population dynamics, plotted in \autoref{fig:time_domain}.b, show the population of state $N_{2}$ gradually increasing until the pulse is switched off. 
It should be noted that state $N_{3}$ rapidly decays non-radiatively into state $N_{2}$ due to the very fast $\tau_{32}$. 
This creates a population inversion between states $N_{2}$ and $N_{1}$, which is probed at $t\approx 10 \unit{ps}$ to produce significant emission.
When the probe reaches the metamaterial, we can observe a reduction in the inversion rate due to generation of stimulated light emission (see ~\autoref{fig:time_domain}.d green line).
To quantify the change in the emission due to the presence of the quantum emitters, we take the difference of the Fourier-transformed fields in the central nanogap with and without the four-level system, plotted in~\autoref{fig:time_domain}.c. 
We see a large emission peak at $f\approx 93 \unit{THz}$ for a 
coupling coefficient $\sigma=21\unit{\Omega^{-1} m^{-1}}$, and absorption peaking (corresponding to negative $\Delta EF$) at frequencies $f>120 \unit{THz}$. 
On the same figure, we also show the normalized emission and absorption linewidths of the four-level system with shaded area. 
The emission peak is shifted from the normalized emission peak, partially due to the non-linearity brought by the four-level system.
Additionally, the anapole is formed as an interference of two eigenmodes and it is near an exceptional point~\cite{canos_valero_theory_2021,zhang_non-hermitian_2025}, and as such, increasing the pumping further causes mode switching of the light emission~\cite{doppler_dynamically_2016,fischer_controlling_2024}, while also increasing intensity and reducing linewidth (see SM S.VIII for further details).
On the other hand, if the coupling coefficient is reduced, then the emission peak is also reduced as expected, since the system absorbs and stores less energy, and also aligns with the emission frequency of the four-level description better. 
In~\autoref{fig:time_domain}.d we show the field in the central gap for various coupling coefficients, in comparison with the incident pump. It is evident that the largest coupling coefficient for the four-level system produces the largest amplification for the light emission when the probe stimulates the emission. 
While the small dipole moment associated with vibrational states of molecules~\cite{kongsuwan_suppressed_2018, chikkaraddy_single-molecule_2023} would make it challenging to achieve lasing in practice, the proposed system is ideal for further enhancing previous experimental realizations of single molecule vibrational spectroscopy~\cite{chikkaraddy_single-molecule_2023}, vibrationally assisted frequency conversion~\cite{xomalis_detecting_2021, chen_continuous-wave_2021} and vibrational pumping~\cite{jakob_optomechanical_2025}.

\section{Conclusion}
By assembling nanoplasmonic aggregates with deeply subwavelength gaps, we have demonstrate high refractive-index metamaterials that support exceptional field enhancement in the MIR regime. 
The resulting aggregates behave as a nearly dispersionless metamaterial across the MIR, while reaching refractive indices of $n>20$ near the near-infrared resonance. 
When shaped into a cylindrical disc, a high-index resonator is formed that supports an anapole state and a near-field that extends across the metamaterial unit cells. 
At the anapole frequency, the local fields confined within the nanoparticle gaps are further enhanced, producing a combined intensity enhancement of up to five orders of magnitude. 
Introducing quantum emitters into these nanogaps enable extreme light-matter interactions with amplified light emission, which establishes self-assembled nanoplasmonic aggregates as a scalable platform for high-index metamaterial with extreme nanoscale field confinement. 
This opens opportunities for enhanced sensing, nonlinear optics, single-molecule vibrational spectroscopy, vibrationally assisted frequency conversion and vibrational pumping.

\section{Methods}
To calculate the effective refractive index of the homogenized material, we use COMSOL Multiphysics (v6.1) RF module. The simulation domain is a hexagonal unit cell for spheres and a square unit cell for cubes, with Floquet periodic boundary conditions applied on the sides and a perfectly matched layer applied at the top and bottom. An incoming port is added at the bottom to launch the excitation and obtain the complex reflection coefficient, and an outgoing port is added at the top to obtain the complex transmission coefficient. The ports are set to be half a wavelength away from the metamaterial layer. From the complex refrlection and transmission coefficient we obtain the effective parameters via standard inversion methods~\cite{chen_robust_2004}. The permittivity of gold is given by a Drude–Lorenz model with set parameters as in~\cite{elliott_fingerprinting_2022}. The scattering and absorption cross section is calculated in a spherical physical domain surrounded by perfectly-matched-layers, using the standard frequency domain scattered field formalism in COMSOL, with a plane wave as excitation. For the inhomogeneous structure the mesh of the nanoparticles is set to a minimum of 10nm to ensure computational feasibility. The time domain calculation with Maxwell-Bloch equations for a four-level system is conducted in COMSOL using the Electromagnetic waves, Transient interface, in a rectangular physical domain with scattering boundary conditions, with one of the boundaries launching a pulse wave as excitation. The time stepping size is set to 0.16fs to ensure numerical stability, resulting in simulation times of approximately 8 days on an Intel Xeon Gold 6154 CPU.

\medskip
\textbf{Supporting Information} \par 
Supporting Information is available at \cite{sztranyovszky_supplementary_2025}.

\medskip
\textbf{Data availability} \par 
The data supporting the findings of this article is available at

\medskip
\textbf{Conflict of interest} \par 
The authors declare no conflict of interest.

\section*{Acknowledgements}
AD acknowledges support from the Royal Society University Research Fellowship
URF\textbackslash R1\textbackslash 180097 and URF\textbackslash R\textbackslash 231024, Royal Society Research Fellows Enhancement Award RGF
\textbackslash EA\textbackslash 181038, funding from ESPRC grants EP/Y008774/1 and EP/X012689/1 from EPSRC for the CDT in
Topological Design EP/S02297X/1. N.S. acknowledges support from EPSRC Grant EP/L015889/1 for the EPSRC Centre for Doctoral Training in Sensor Technologies and Applications, and from AstraZeneca (MedImmune Ltd). C.T. is supported by a Gates Cambridge fellowship (OPP1144). R.A. acknowledges support from the Winton Programme for the Physics of Sustainability and from St. John’s College Cambridge. RC acknowledge funding from UKRI Future Leaders Fellowship.


\printbibliography[heading=subbibliography,title={References}]
\end{refsection}

\newpage
\renewcommand{\thesection}{S \Roman{section}} 
\renewcommand{\thesubsection}{\thesection.\Roman{subsection}}
\begin{refsection}
\setcounter{section}{0}
\setcounter{figure}{0}
\begin{titlepage}
    \centering
    {\Huge Supplementary Material}
\end{titlepage}

\section{Spectra of single gold sphere and infinite metasurface}
\label{s:modes}

\begin{figure}[h]
    \centering
    \includegraphics[width=0.7\linewidth]{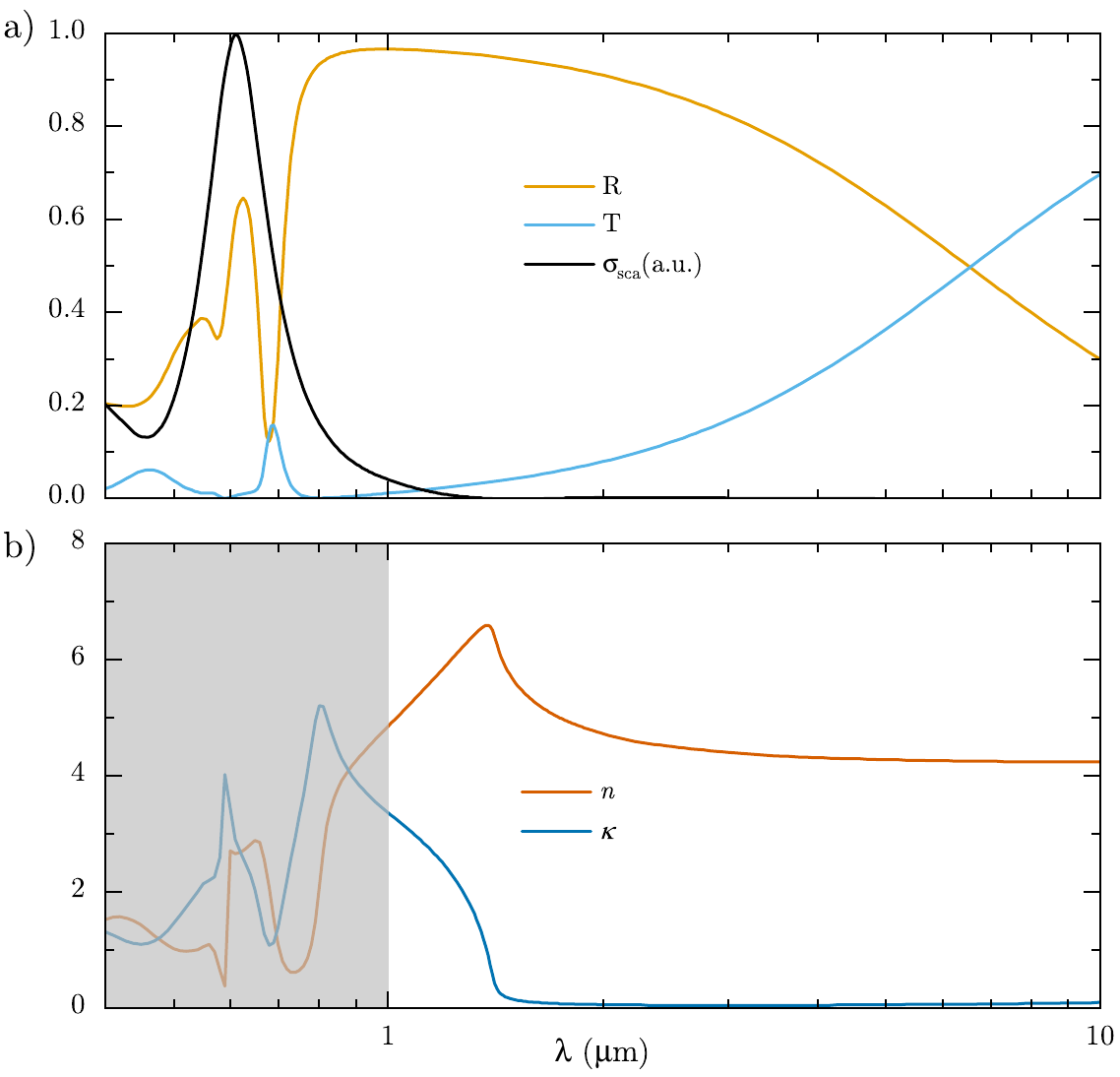}
    \caption{a) Reflection $R$ and transmission $T$ spectra for a MM made of gold nanospheres of diameter $d=100$nm, on a hexagonal lattice, with $g=1$nm gap size between them, embedded in a homogeneous slab of thickness $d$ and refractive index $n_\textrm{b}=1.5$, surrounded by air, as illustrated on Figure 1.a of the main text. Scattering cross section $\sigma_\textrm{sca}$ of a single gold sphere with diameter $d$, in air. b) Effective parameters for the system, refractive index $n$ and loss $\kappa$. Shaded area indicates the region where the effective parameters might not be valid.}
    \label{fig:FP_and_SP}
\end{figure}

On \autoref{fig:FP_and_SP}.a we show reflection $R$ and transmission $T$ spectra of a metamaterial (MM) made of a single infinite layer of gold nanospheres on a hexagonal lattice with diameter $d=100$nm and gap size $g=1$nmm, embedded in a slab of refractive index $n_\text{b}=1.5$ and thicknes $d$, surrounded by air, alongside with the scattering cross section spectrum $\sigma_\textrm{sca}$ of a single gold nanosphere with the same diameter, in air. The scattering cross section shows the well know surface plasmon (SP) resonance in the visible range~\cite{seehra_noble_2018}. For the MM, in the visible region the effective medium approximation (EMA) might not hold due to the short wavelength with respect to the size of the structure, and we can attribute the resonances to collective resonances originating from the coupled modes of the nanospheres. In particular, the resonance in the reflection spectrum aligns well with the SP mode of the single sphere at $\lambda\approx610$nm. At longer wavelengths we can assume the EMA to be valid, and thus we treat the MM as a homogeneous slab. A homogeneous slab has Fabry-P\'erot (FP) modes at $\lambda_m=2Ln_s/m$, where $L$ is the thickness of the slab, $n_\text{s}$ is the refractive index, and $m$ is an integer. Thus, the increase in transmission at longer wavelengths is due to the $m=0$ FP mode at wavenumber $k=0$~\cite{doost2012resonant}, \textit{i.e.} at $\lambda=\infty$. \autoref{fig:FP_and_SP}.b shows the corresponding effective real refractive index $n$ and effective losses $\kappa$. The shaded area indicates the region where the EMA might not be valid. Following from the Maxwell Garnet theory, for the EMA to hold it is typically required that $ka<<1$, where $k$ is the wavenumber of light and $a$ is the characteristic size of the meta-atoms~\cite{mallet_maxwell-garnett_2005,aspnes_plasmonics_2011}. This leads to the well known rule-of-thumb that the wavelength should be five to ten times longer than the characteristic size of the meta-atoms, which in this case, means longer than $\lambda=1$\textmu m.  

\newpage
\section{Induced current in spherical meta-atoms}
\label{s:current_sphere}

\begin{figure}[h]
    \centering
    \includegraphics[width=0.9\linewidth]{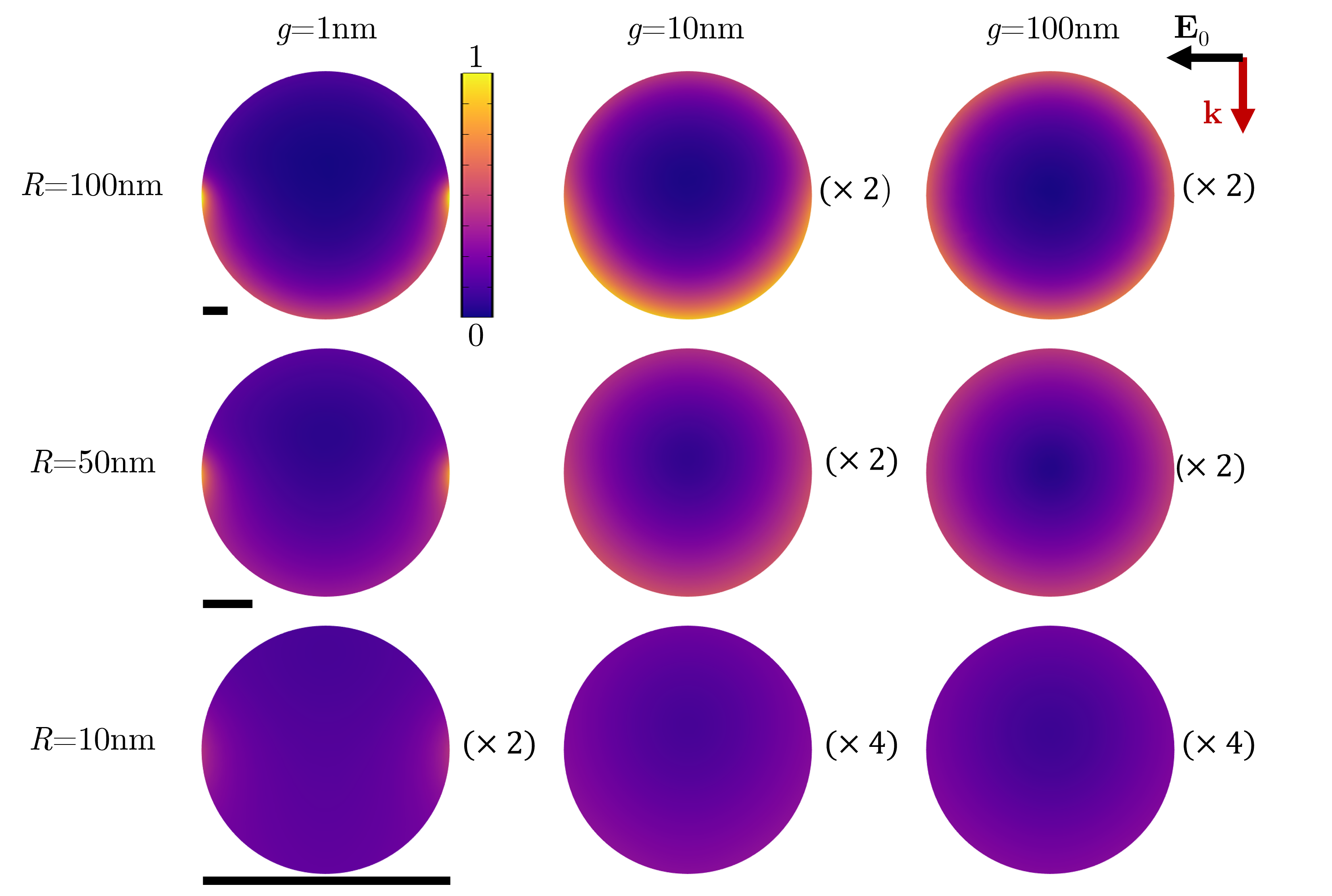}
    \caption{Absolute normalized current density $|\vb J|$ in the nanospheres of the MM at $\lambda=10$\textmu m. The number in the brackets indicate by how much the values were scaled up for better visibility. The black line on the left is a scale bar indicating 20nm for reference, and the inset on the top right indicates the incoming electric field $\vb E_0$ orientation and the wavevector $\vb k$.}
    \label{fig:current}
\end{figure}

On \autoref{fig:current} we show the induced current distrubution inside the gold nanoparticles of the MM, for different particle radiuses $R$ and different gap sizes $g$. In all cases, the current is stronger at the surface, and decays towards the center of the particle, as expected from the skin depth of the metal. For large gaps the current distribution is similar to that of single particles. For decreasing gap size hotspots emerge near the small gaps where the field is concentrated due to the coupling with neighboring particles, and the current flow is shifted towards the bottom of the particles. For decreasing particle size the current is significantly reduced due to the smaller volume of free electrons and decreasing coupling between neighbors.

\newpage
\section{Imaginary part of $\varepsilon$ and $\mu$}
\label{s:imaginary_parts}

\begin{figure}[h]
    \centering
    \includegraphics[width=0.9\linewidth]{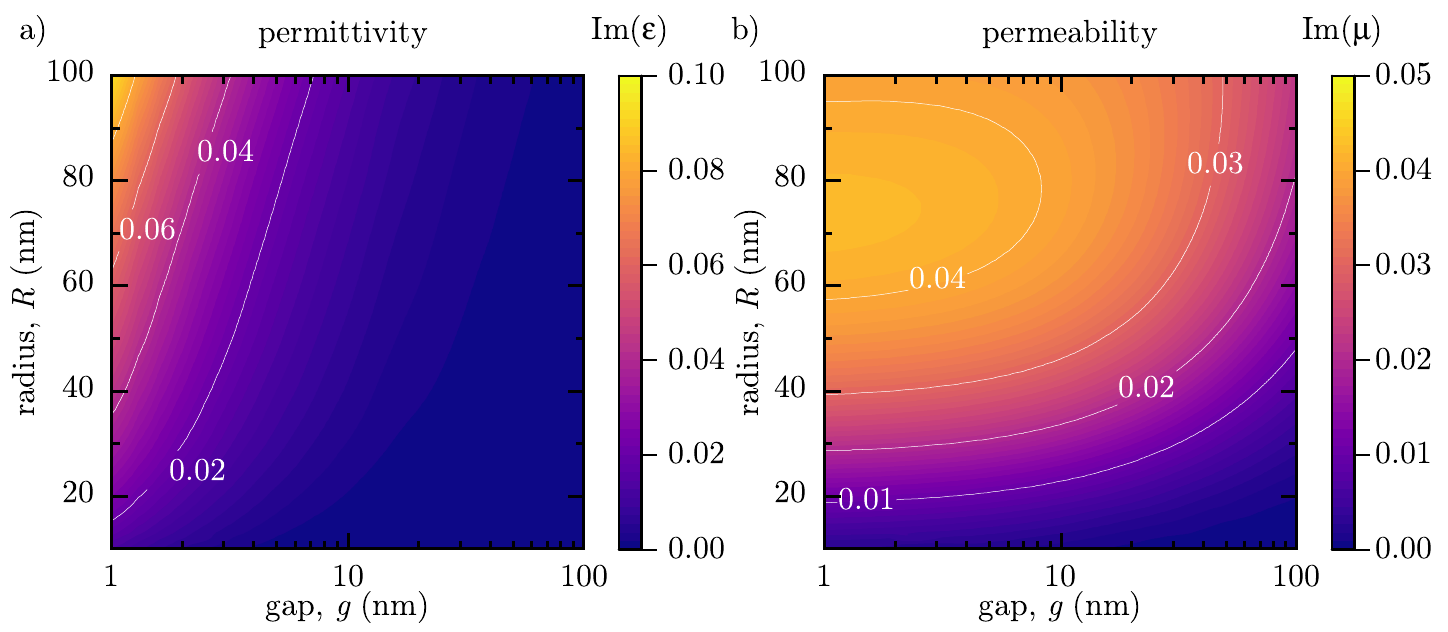}
    \caption{Imaginary part of the permittivity and permeability for a MM composed of a single layer of gold spheres of radius $R$, with interparticle gap size $g$, at $\lambda=10$\textmu m.}
    \label{fig:imaginary_parts}
\end{figure}

On \autoref{fig:imaginary_parts} we show the imaginary parts of the effective permittivity and permeability, indicating the electric and magnetic losses, as a function of radius $r$ and gap size $g$, at long wavelength $\lambda=10$\textmu m. Consistently with the field enhancement and the real part of the permittivity (shown in Figure 1.e of the main text), the imaginary part $\text{Im}(\varepsilon)$ is increasing for decreasing $g$ and increasing $r$. On the other hand, while the real part of the permeability (shown in Figure 1.f of the main text) is decreasing for decreasing $g$ and increasing $r$, the imaginary part $\text{Im}(\mu)$ is increasing. Both are the consequence of the increasing current: while $\text{Re}(\mu)$ is decreasing due to the increasing induced opposing magnetic field, in the EMA the overall losses due to the current flow, which are increasing with the field enhancement, are distributed between effective electric and magnetic losses, thus $\text{Im}(\mu)$ is increasing. 

We can write the complex refractive index as 
\begin{align}
\underline{n} = \sqrt{\varepsilon\mu}=\sqrt{\Re(\varepsilon)\Re(\mu)+i[\Re(\varepsilon)\Im(\mu)+\Im(\varepsilon)\Re(\mu)]-\Im(\varepsilon)\Im(\mu)} \,.
\end{align}
Noting that $\text{Re}(\varepsilon)\gg\text{Im}(\varepsilon)$ and $\text{Re}(\mu)\gg\text{Im}(\mu)$, we can drop the last term $\Im(\varepsilon)\Im(\mu)$ inside the square root. Then, we can apply a Taylor expansion around the real part of the expression and write the complex refractive index as 
\begin{align}
\underline{n} \approx \sqrt{\Re(\varepsilon)\Re(\mu)} + \frac{1}{2} \frac{i[\Re(\varepsilon)\Im(\mu)+\Im(\varepsilon)\Re(\mu)]
}{\sqrt{\Re(\varepsilon)\Re(\mu)+i[\Re(\varepsilon)\Im(\mu)+\Im(\varepsilon)\Re(\mu)]}} \,. \end{align}
Neglecting the contribution of  the imaginary parts in the denominator, then taking the real and imaginary part of the complex refractive index we arrive at 
\begin{align}
n&\approx\sqrt{\text{Re}(\varepsilon)\text{Re}(\mu)} \,, \\
\kappa &\approx \left[\text{Re}(\varepsilon)\text{Im}(\mu) + \text{Im}(\varepsilon)\text{Re}(\mu)\right]/2/n \,,
\end{align}
which shows that we can neglect the contribution of $\text{Im}(\varepsilon)$ and $\text{Im}(\mu)$ to the real refractive index $n$, but they are important to interpret the changes in the loss rate $\kappa$.

\newpage
\section{Change of host refractive index}
\label{s:host_index}

\begin{figure}[h]
    \centering
    \includegraphics[width=0.9\linewidth]{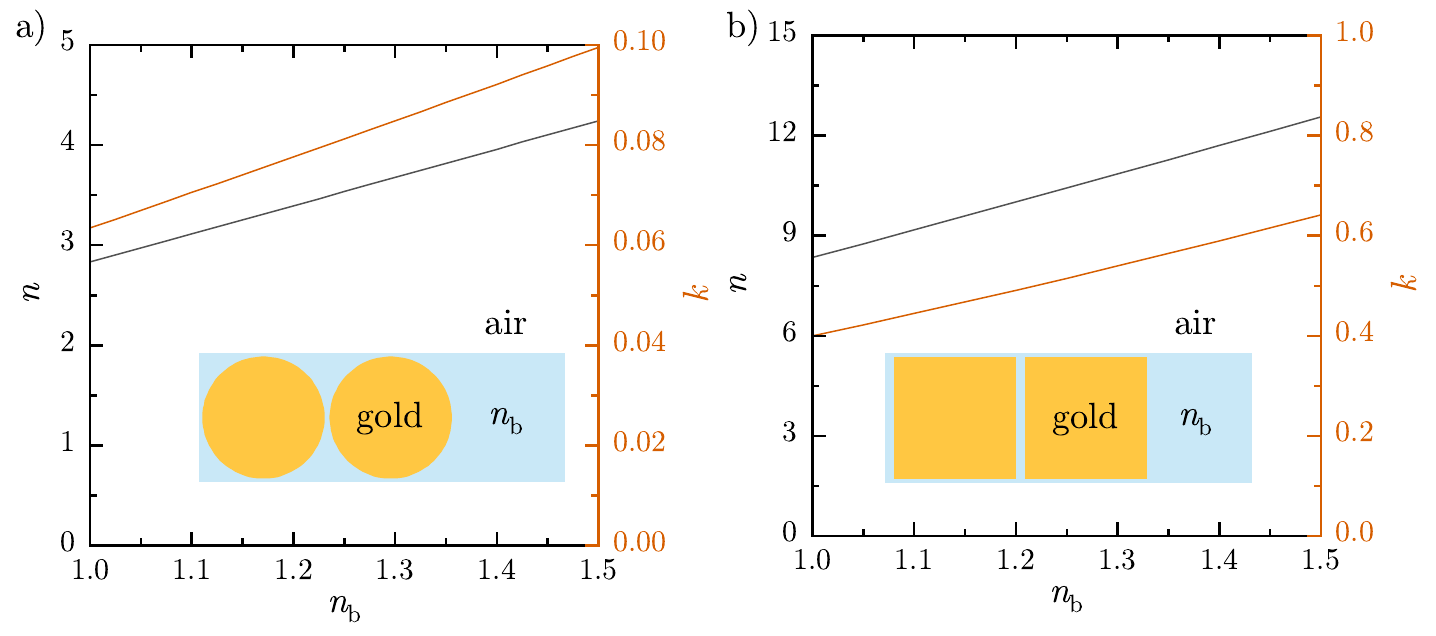}
    \caption{Effective refractive index as a function of host refractive index at $\lambda=10$\textmu m, for a MM made of a single layer of: a) gold spheres with diameter of 100nm on a hexagonal grid with 1nm gaps; b) gold cubes of edge length 100nm on a square grid with 1nm gaps and 1nm edge rounding.}
    \label{fig:host_index}
\end{figure}

On \autoref{fig:host_index} we show how the effective refractive index $\underline{n}$ changes as the refractive index of the host medium $n_\text{b}$ changes, for a gold sphere and a gold cube. In both geometries we can observed that $\underline{n} \propto n_\text{b}$, with both the real part $n$ and imaginary part $\kappa$ scaling linearly with the host refractive index. These results are consistent with the Maxwell Garnet mixing formula~\cite{garnett_colours_1904}, which can be wirtten in the approximate form as $n\approx n_\text{b} \sqrt{(1+2f)/(1-f)}$, where $f$ is volume fraction of gold in the unit cell~\cite{spiesshofer_tailoring_2025}. 

\newpage
\section{Number of layers}
\label{s:layers}

\begin{figure}[h]
    \centering
    \includegraphics[width=0.9\linewidth]{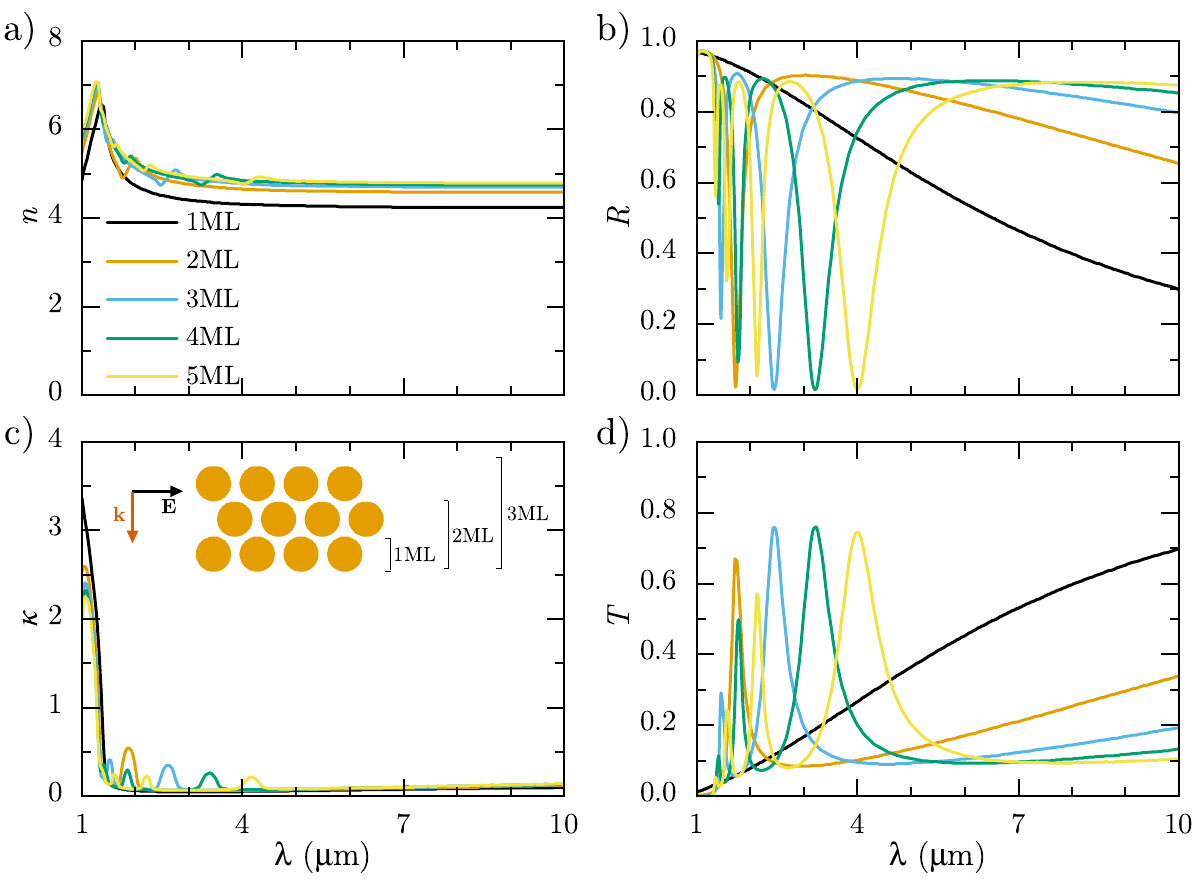}
    \caption{Effective refractive index for a MM made of gold spheres with $d=100$nm, and $g=1$nm, on a closely packed hexagonal lattice, for different number of layers. a) Real part of the effective refractive index. b) Reflection spectra. c) Imaginary part of the effective index. d) Transmission spectra.}
    \label{fig:layers}
\end{figure}

On \autoref{fig:layers}.a  and \autoref{fig:layers}.c we show the real part $n$ and imaginary part $\kappa$ of the effective refractive index, respectively, when the number of layers is the MM is varied. With the increasing number of layers the volume fraction $f$ of gold in the unit cell rapidly increases from $f\approx0.6$ for a single layer, to the theoretical maximum of $f\approx0.7$ for a multilayer on a closely packed hexagonal lattice, with $d=100$nm and $g=1$nm. In accordance with this, the effective refractive index also increases. With the increasing number of layers we can observe small oscillations appear in the range $1.5$\textmu m $<\lambda<4$\textmu m, which correspond to FP modes of the slab, see  \autoref{fig:layers}.b for the reflection spectrum and \autoref{fig:layers}.d for the transmission spectrum. We can observed new modes forming in the mid-IR due to the thickness of the slab increasing.

\newpage
\section{Shape effects}
\label{s:shapes}

In this section we look at how the shape of the meta-atom affects the response of the system. In \autoref{ss:r_and_t} we present the transmission and reflection spectra for faceted spheres and cuboid particles, in \autoref{ss:eps_and_mu} we show the effective permittivity and permeability for the same systems, and in \autoref{ss:edge_rounding} we look at the effect of the edge rounding on the effective refractive index in case of a cube meta-atom.

\subsection{Reflection and transmission}
\label{ss:r_and_t}

\begin{figure}[hb]
    \centering
    \includegraphics[width=0.9\linewidth]{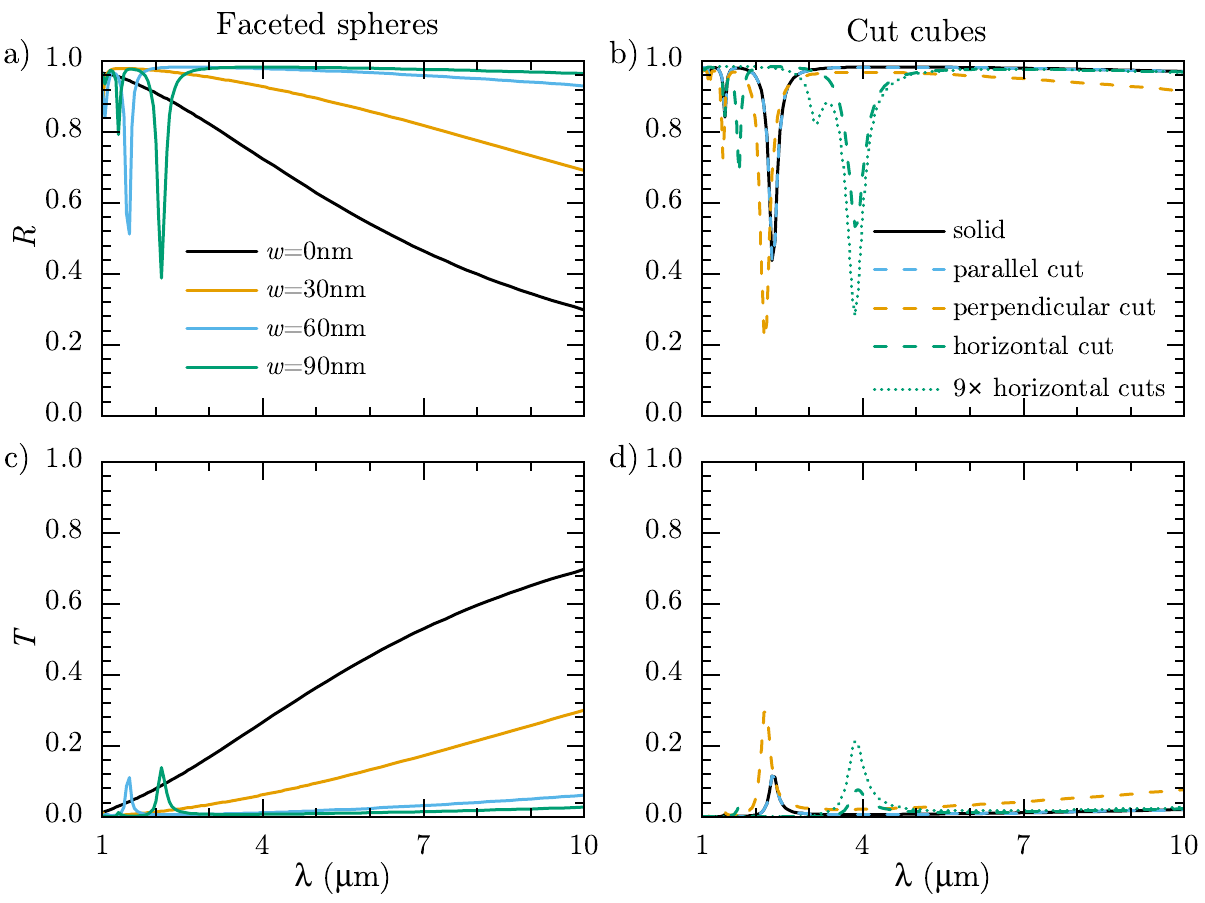}
    \caption{a) Reflection spectra of a MM composed of faceted spheres with diameter $d=100$nm, gap size $g=1$nm, and facet size $w$ on a hexagonal lattice. b) Reflection spectra of a MM composed of cuboid nanoparticles with a characteristic size of 100nm, gap size $g=1$nm, and cuts introduced to them as defined in the main text. c) As a) but transmission spectrum. d) As b) but transmission spectrum.}
    \label{fig:shapes_RT}
\end{figure}

On \autoref{fig:shapes_RT}.a and \autoref{fig:shapes_RT}.c we show the reflection and transmission spectra, respectively, for a MM composed of a single layer of faceted spheres. As shown in the main text, increasing the faceting of the spheres results in an increasing refractive index, and accordingly, the spectra redshifts. For a small facet of $w=30$nm we can observe the $m=1$ FP mode as a dip in the reflection spectrum at $\lambda=1$\textmu m. With a larger facet the spectra further redshifts, and the $m=2$ mode also becomes visible. 

\begin{figure}[h]
    \centering
    \includegraphics[width=0.8\linewidth]{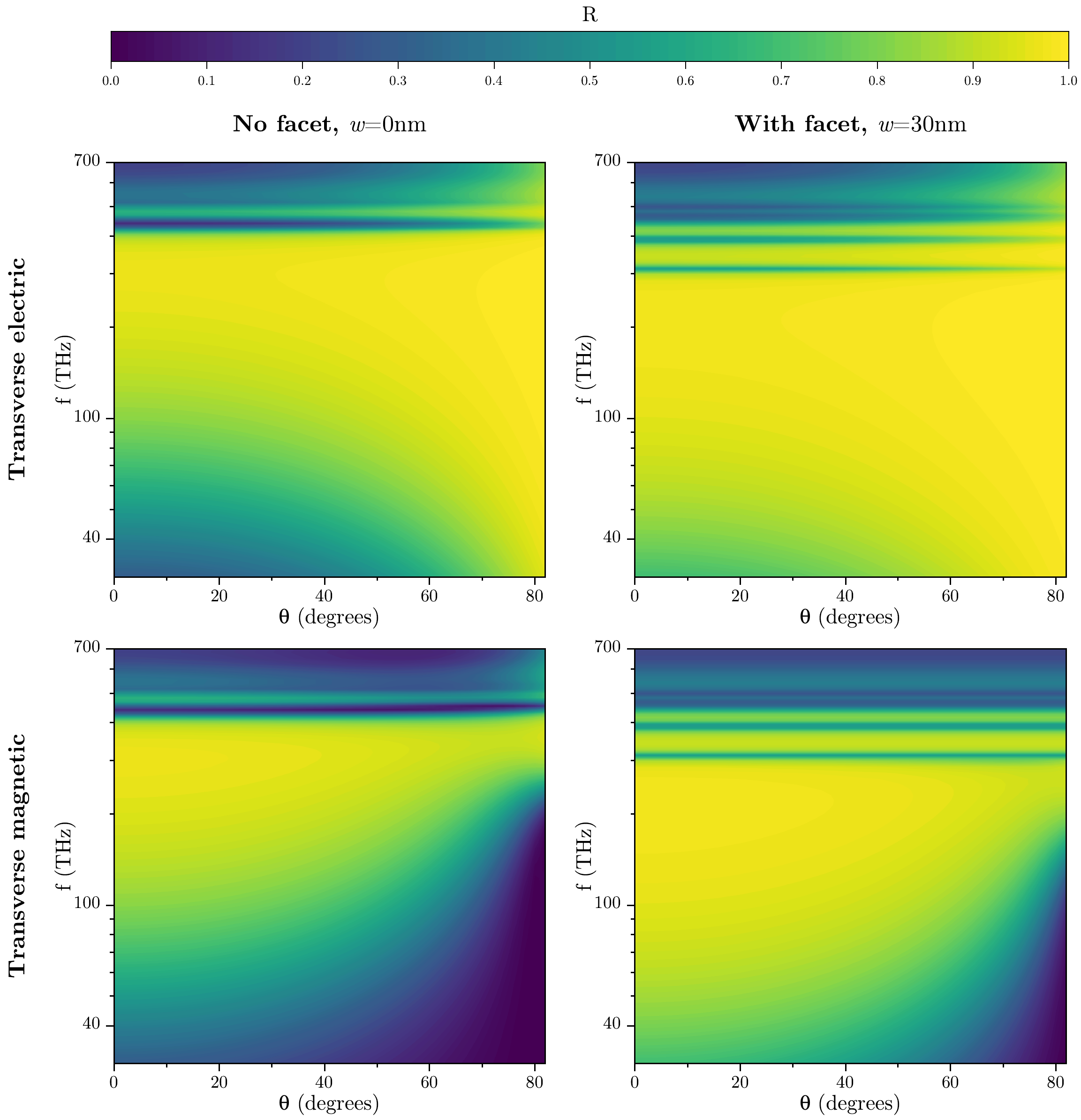}
    \caption{Angular dependent refection spectra for a single layer metamaterial made of spherical nanoparticles, without faceting (left column) and with a $w=30$nm facet (right column), for transverse electric (top row) and transverse magnetic (bottom row) polarizations.}
    \label{fig:bands}
\end{figure}

In accordance with the FP modes redshifting, with the increasing refractive index the effective photonic band also reaches the Brilluin zone at lower frequencies, see \autoref{fig:bands}. In the angular dependent reflection spectra we can see that the edge of the reflection band shifts starts at lower frequencies for both transverse electric and transverse magnetic polarization, across all angles. For low frequencies ($f<200$THz) in the transverse magnetic polarization we can see a reduction in reflection for higher angles as opposed to the increase in transverse electric polarization, which is due to Brewster's angle. For angles beyond $\theta=85^\circ$ the reflection increases for transverse magnetic polarization as well, however due to issues with numerical stability in the simulation this region is not shown.

On \autoref{fig:shapes_RT}.b and \autoref{fig:shapes_RT}.d we show the reflection and transmission spectra, respectively, for a cube that is cut along horizontal and vertical directions, as described in the main text. Due to the high refractive index the first ($m=1$) FP mode is well into the near-IR range, at $\lambda\approx2300$nm. Introducing a parallel cut does not change the spectra compared to the solid cube, while introducing a perpendicular cut results only in a small blueshift of the FP mode, in accordance with only a small change of the refractive at this particular wavelength, see Figure 2.b of the main text. For a horizontal cut the FP modes shifts to longer wavelength due to the refractive index having a large peak between $3$\textmu m $<\lambda<4$\textmu m. Interestingly, while for more horizontal cuts the real refractive index increases further, the mode does not redshifts further. This could possibly be due to the imaginary part of the refractive index significantly decreasing, which may also influence the position of the modes.

\newpage
\subsection{Permittivity and permeability}
\label{ss:eps_and_mu}
\begin{figure}[h!]
    \centering
    \includegraphics[width=0.8\linewidth]{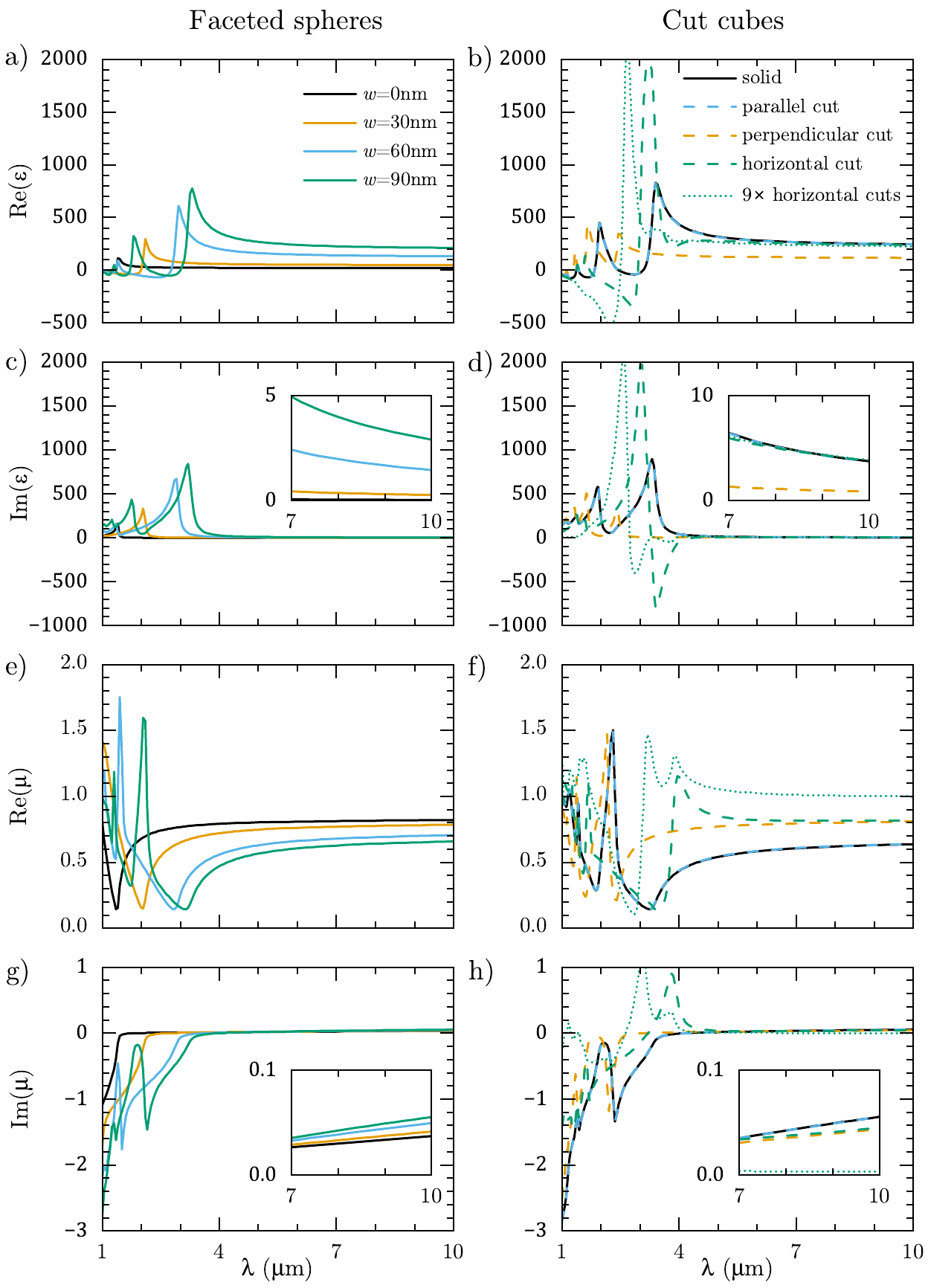}
    \caption{Effective parameters for metasurfaces made of faceted spheres of diameter $d=100$nm and gap size $g=1$nm (left column), and cuboidal nanoparticles with characteristic size of 100nm and gap size $g=1$nm (right column). First row is the real part of the permittivity, the second row is the imaginary part of the permittivity, the third row is the real part of the permeability, and the fourth row is the imaginary part of the permeability.}
    \label{fig:shapes_eps_mu}
\end{figure}

\begin{figure}
    \centering
    \includegraphics[width=0.5\linewidth]{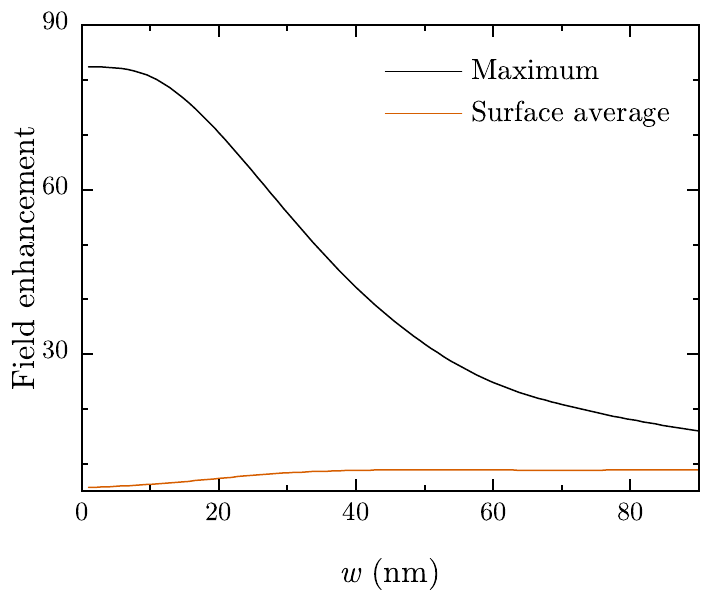}
    \caption{Field enhancement at the center of the gap as a function of facet size. Top and bottom facets, where there are no gaps, are excluded when calculating the surface average.}
    \label{fig:facet_field}
\end{figure}

On \autoref{fig:shapes_eps_mu} we show the real and imaginary part of permittivity and permeability. To describe the changes in the systems we focus on the long wavelength behavior for simplicity. First considering the spheres, introducing faceting increases the amount of gold, thus the polarizability in the unit cells, resulting in an increased permittivity. We note that this is despite the maximum field enhancmeent at the center of the gap decreasing, albeit the surface averaged field enhancement increases (see \autoref{fig:facet_field}).  With increasing facets the effective size of the particle increases, resulting in more current (see \autoref{s:current_sphere}), a larger induced opposing magnetic field, thus a decrease in permeability. In accordance with this, both magnetic and electric losses are increasing at long wavelength. 


Fundamentally, the dielectric behavior of the MM emerges from the gaps in the metallic structure, effectively impeding the free flow of the electrons and binding them to the meta-atoms. Therefore, instead of simply optimizing the size of the cube for maximum achievable refractive index, we now change its polarizability and the induced current flow by introducing new cuts to the NP. This will change the permittivity and permeability, and thus the refractive index. In effect, the cubic meta-atom in the unit cell will be made from smaller rectangular cuboids. We consider three different configuration: for \textit{parallel cut} a cut plane is introduced into the cube that is parallel to the incident field $\vb E_\textrm{0}$ and to the wavevector $\vb k$ as well; for \textit{perpendicular cut} the cut plane is perpendicular to $\vb E_\textrm{0}$ and parallel to $\vb k$; for \textit{horizontal cut} the cut plane is parallel to $\vb E_\textrm{0}$ and normal to $ \vb k$. In all cases the cut introduces a 1nm gap into the structure, the same gap size as between the meta-atoms in each unit cell. We note that introducing cuts can results in increasing non-locality and anisotropy. 

\begin{figure}[b]
    \centering
    \includegraphics[width=0.99\linewidth]{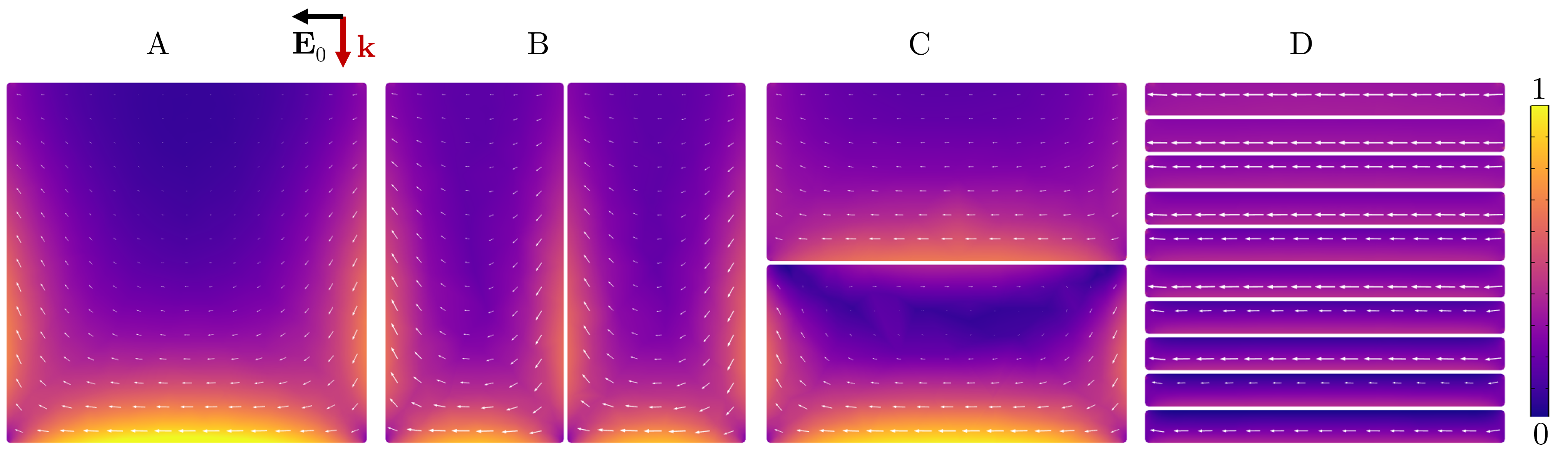}
    \caption{Normalized current distribution $|\vb J|$ inside the cuboidal meta-atoms. `A' is the solid cube without a cut, `B' has a perpendicular cut, `C'  has the horizontal cut, and `D' has 9 horizontal cuts. The incoming field $\vb E_0 $ and propagation direction $\vb k$ is indicated by the arrows.}
    \label{fig:current_cubes}
\end{figure}

Considering the parallel cut, this does not change the refractive index. We can understand this result based on the contribution of the induced dipole moment $\vb p$ to the polarizability $\alpha=|\vb p|/|\vb E_0|$, as well as the induced current flow. For a solid cube we take the approximation that we have one dipole in the center of the unit cell, with induced charge $q$ and displacement $\vb d$, resulting in the induced dipole moment $\vb p=q\vb d$. After introducing the parallel cut we have two dipoles, each with half the induced charge, but the original displacement available, therefore, there is no change in the overall polarizability. In addition, due to their alignment with each other and the incoming field, the dipoles are always excited together in-phase. Furthermore, the induced current flow (see \autoref{fig:current_cubes}) loops in the plane parallel to both $\vb E_\textrm{0}$ and $\vb k$, and there is no change in the morphology in this direction, therefore, the loss and the permeability remains unaffected, see \autoref{fig:shapes_eps_mu}. 

When we introduce a perpendicular cut, apart from the region $\lambda<1.5$\textmu m where $n$ is mostly unchanged, the refractive index and losses decrease. In the dipole approximation now we replace the one dipole with two smaller ones, where both the charge and the displacement are reduced by a factor of two. Overall, this would result in a polarizability reduced by a factor of 2, and the refractive index reduced by a factor of $\sqrt 2$, i.e. by about $30\%$. For the permittivity this is true at long wavelengths, and it is approximately in line with the change in the refractive index, which reduces from $n\approx12.55$ by about $20\%$ to $n\approx9.8$, instead of the expected $n\approx8.75$. The difference is coming from that now the permeability has increased from $\mu=0.637$ by nearly $30\%$ to $\mu=0.816$. The increase in the permeability is inline with the reducing loss, they are anti-correlated, as observed in Figue 1 of the main text. Both can be understood from the perspective of induced current flow, where the loop is now impeded along the direction of $\vb E_0$ by the cut introduced perpendicular to it. At short wavelength, while both the permeability and permittivity changes significantly they do so in opposite ways, thus the overall effect leaves the refractive index mostly unchanged. Note that in this configuration, both cuboids are still excited in-phase with each other. 
 
Introducing a horizontal cut increases the refractive index of the low frequency resonance (at $\lambda=3.04$\textmu m), and moderately in the long wavelength limit, while it can decreases it for shorter wavelengths ($\lambda<2.5$\textmu m). On resonance the loss increases significantly, while for long wavelength it reduces. In the dipole approximation once again we have two dipoles, with half the charge but the original displacement available, thus one would expect the same polarizability. This is indeed the case in the long wavelength limit, however the two dipoles are not excited perfectly in-phase, and this gives rise to significantly different behavior in the different frequency regions, depending on which coupled modes are excited. For long wavelengths the current loop is inhibited similarly to the perpendicular cut case, except now along the direction of $\vb k$, resulting in reduced losses, increased permeability, and the overall increased refractive index. In the region $3$\textmu m$<\lambda<4$\textmu m, both $\mu$ and $\varepsilon$ exhibit strong resonances, giving rise to the two peaks in the refractive index, and the increased losses. For shorter wavelength the refractive index can be smaller or the same as it was for the solid cube, depending on the phase of the two coupled dipoles.

Based on these results we can expect an even higher refractive index if we slices the cube into more and thinner horizontal slices. Therefore, we introduce nine cuts, creating all together ten thin slices of gold cuboids, each with a height of $h=9.1$nm, stacked with 1nm gaps in between them, giving the original thickness $d=100$nm. Despite nearly $10\%$ decrease in the volume fraction of gold in the unit cell, which does manifest in the decrease of the permittivity, the refractive index generally increases as expected, on resonance (at $\lambda=3.04$\textmu m) and in the long wavelength limit as well, by about $20\%$. This is due to the new cuts further inhibiting the induced current flow to loop around in the direction of $\vb k$, effectively thinning the meta-atom and suppressing the induction of opposing magnetic field. This results in $\mu\approx1$, and overcompensates for the decrease in permittivity. In addition, consistently with the increase of the permeability, the losses are reduced for long wavelengths. The refractive index peaks at an exceptionally high value of $n\approx24.9$, and approximately $20<n$ over a broad range ($2.3$\textmu m$<\lambda<3.6$\textmu m).

We note that for $\lambda<4$\textmu m either the imaginary part of the effective permittivity or the permeability can take negative values in the systems considered here, which corresponds to electric or magnetic gain. On their own, independently, they would be unphysical for a passive material, and this could question the validity of the effective medium description~\cite{simovski2009material}. It is important to note however, that here an electric (magnetic) gain is always accompanied by a magnetic (electric) loss, and the combined value of them in $\kappa$ results in an overall loss, which is physical. Hence, we can interpret the electric (magnetic) gain here as simply shifting the energy stored in the electric (magnetic) field to energy stored in the magnetic (electric) field. In a more strict interpretation, this sets the limit of validity for the EMA. For the perfect spheres the permeability turns to gain at $\lambda\approx1.8$\textmu m where $n\approx4.9$, while for solid cubes this occurs at $\lambda\approx4.2$\textmu m where $n\approx13.5$. These give wavelengths in the medium as $\lambda/n\approx367$nm and $\lambda/n\approx311$nm, for the sphere and the cube, respectively. The limits that we obtained here based on the imaginary part of the permittivity or permeability are more relaxed than the rule-of-thumb that the wavelength should be five to ten times longer than the characteristic size of the meta-atoms. A possible approach to circumvent the electric or magnetic gain, is to take the approximation that the dispersion of refractive index is due to the permittivity only, and assume a non-magnetic material. This will result in a less accurate reproduction of observables, such as transmission or reflection through the MM, however ensures that the material passivity requirement is fulfilled for the permittivity and permeability as well.




\newpage
\subsection{Edge rounding}
\label{ss:edge_rounding}
Here we consider the effect of edge rounding of cubical meta-atoms on the effective refractive index at long wavelengths, far from resonances. For an increasing edge rounding the cube becomes more spherical, the volume fraction of gold in the unit cell decreases, and the refractive index decreases linearly with $r$. In effect, the increasing edge rounding is the opposite of increasing faceting.

\begin{figure}[h]
    \centering
    \includegraphics[width=0.5\linewidth]{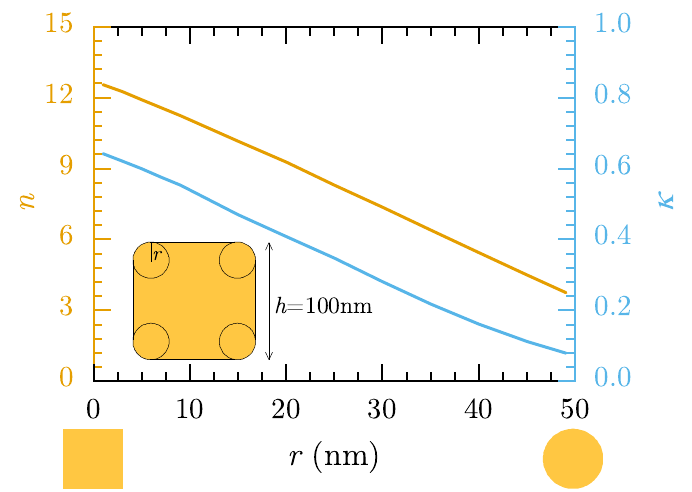}
    \caption{Change of the effective refractive index as the edge rounding $r$ is changed, at $\lambda=10$\textmu m, for a cuboidal meta atom of characetristic size $h=100$nm. Note that $r=0$nm corresponds to a cube with sharp edges, and $r=50$ nm corresponds to spheres on a square grid.}
    \label{fig:edge_rounding}
\end{figure}

\newpage
\section{Anapole}
\label{s:anapole}
In this section we look at how certain parameters influence the scattering and absorption spectra of a dielectric cylindrical MM resonator. In \autoref{ss:anapole_abs} we consider the effect of the effective loss rate of the MM to explain the mismatch between the spectra of the inhomogeneous and homogenous cylindrical structures, and in \autoref{ss:anapole_disp} we show how the anapole position can be tailored by changing the radius or height of the cylindrical resonator.

\subsection{Absorption and field enhancement}
\label{ss:anapole_abs}
\begin{figure}[h!]
    \centering
    \includegraphics[width=0.48\linewidth]{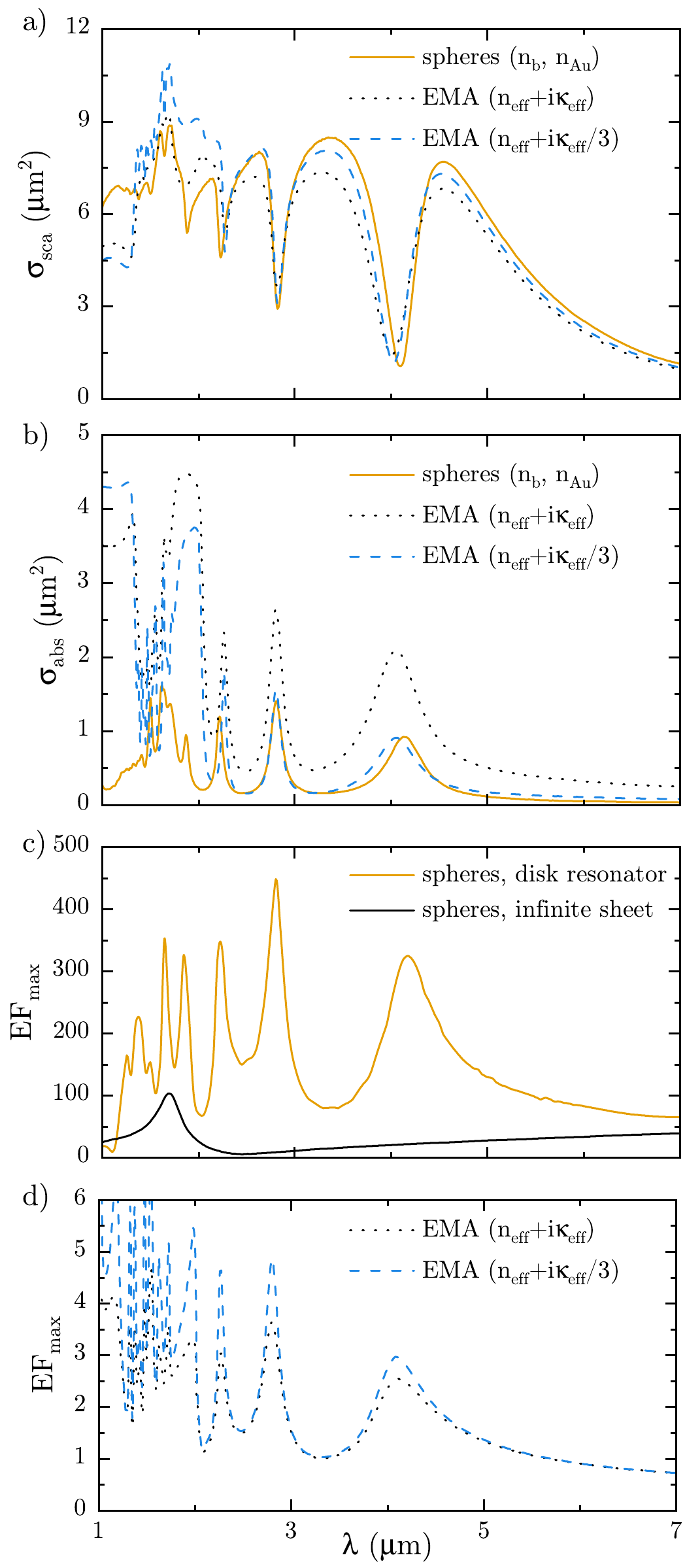}
    \caption{a) Replication of Figure 3.b from the main text. Scattering cross section of a cylinder of radius $R=1$\textmu m and height $h=184$nm, corresponding to two layers of gold spheres. Solid gold line is the inhomogenous structure composed of gold spheres, dotted black line is a cylindrical resonator with effective parameters describing the two layer infinite gold MM, and dashed blue line is the same with losses reduced by a factor of 3. b) as a) but absorption spectra. c) Maximum enhancement factor the disk resonator made of two layers gold spheres (gold line), and in an infinite sheet made of two layers of gold spheres (black line). d) Maximum enhancement factor in a homogenized disk resonator, modelled with effective parameters, with full loss (black dotted line) and when the loss is reduced by a factor of 3 (blue dashed line). }
    \label{fig:anapole_abs}
\end{figure}

On \autoref{fig:anapole_abs}.a we replicate Figure 3.b of the main text, showing the scattering cross section spectra $\sigma_\text{sca}$ for a cylindrical resonator made of gold spheres (gold solid line), the corresponding homogenized cylindrical resonator described with the effective refractive index $\underline{n}=n+i\kappa$, and the same homogenized cylindrical resonator with the material losses reduced by a factor of 3 ($\kappa \rightarrow \kappa/3$). On \autoref{fig:anapole_abs}.b we show the absorption cross section spectra $\sigma_\text{abs}$ for the same configurations. The peaks in absorption at $\lambda\approx4$\textmu m correspond to the anapole, and rise due to the enhanced field inside the lossy structure, see \autoref{fig:anapole_abs}.c. Based on the dips in $\sigma_\text{sca}$, we can attribute the peaks at $\lambda\approx2.8$\textmu m,  $\lambda\approx2.2$\textmu m, and  $\lambda\approx1.9$\textmu m to higher order anapoles~\cite{totero_gongora_fundamental_2017}. On the absorption spectra we can see that with the initially retrieved effective parameters the absorption of the homogenized disk (black dotted line) is significantly higher across the whole IR range than the absorption of the inhomogeneous structure made of gold spheres. This suggest that the effective parameter retrieved for the infinite metasurface overestimates the loss for a finite cylindrical MM resonator. It is known that the shape of the object can impact the effective parameters~\cite{guerra_effective_2025}. Reducing the loss rate by a factor of 3 reduced the difference significantly (blue dashed line), and makes the absorption spectra of the homogenized and inhomogeneous structure match well in the range $\lambda>2.5$\textmu m. 

On \autoref{fig:anapole_abs}.c we compare the field enhancement factor (EF) in the cylindrical resonator made of gold spheres (gold solid line) and for an infinite sheet of gold spheres (black solid line). We can see multiple sharp peaks for the cylindrical resonator, corresponding the fundamental and higher order anapoles, exceeding 300. For an infinite sheet made of two layers of spheres we can see one peak at $\lambda=1.65$\textmu m corresponding to the $n=1$ FP mode, and increasing EF at long wavelength corresponding to the $n=0$ FP mode at $k=0$. The disk resonator provides a significantly higher EF compared to the infinite sheet, in particular about a factor of 14 higher at the fundamental anapole $\lambda\approx4$\textmu m. This is due to the more efficient light coupling with the modes of the disk resonator~\cite{xomalis_interfering_2021}. On \autoref{fig:anapole_abs}.d we show the EF for the homogenized disk resonator. The spectral position of peaks beyond $\lambda=2$\textmu m show a good agreement with the inhomogeneous disk resonator, with naturally a lower total EF. Decreasing the loss rate increases the EF, as less field is absorbed by the structure.

\newpage

\newpage
\subsection{Dispersion}
\label{ss:anapole_disp}

\begin{figure}[h]
    \centering
    \includegraphics[width=0.9\linewidth]{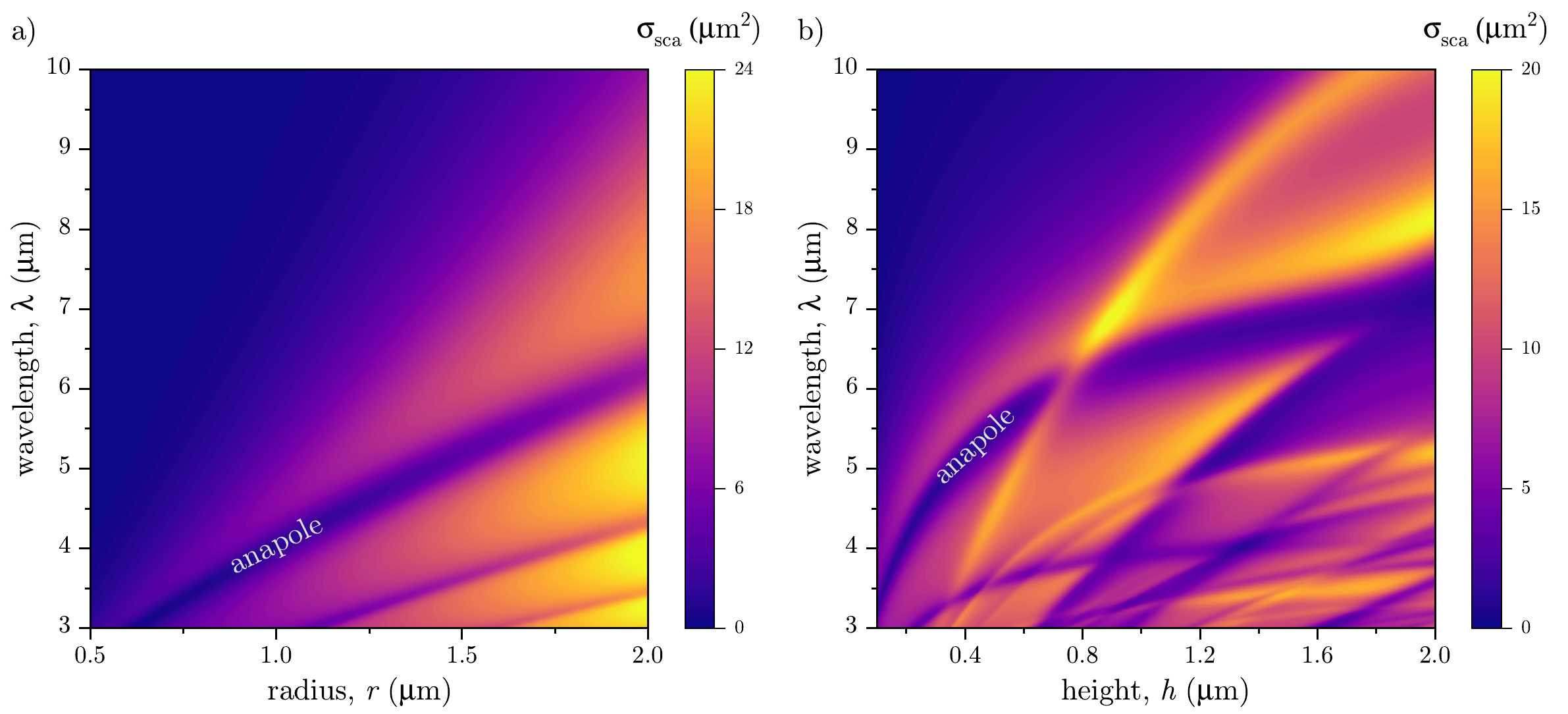}
    \caption{Dispersion diagrams of a cylindrical metamaterial resonator of height $h$ and radius $r$, composed of gold nanospheres with diameter $d=100$nm, interparticle gap size $g=1$nm and host refractive index $n_\text{b}=1.5$. a) scattering cross section spectrum for changing height. b) scattering cross section spectrum for changing radius $r$. }
    \label{fig:anapole_disp}
\end{figure}

One easy way to control the anapole position is via the size of the resonator. On \autoref{fig:anapole_disp}.a  we show how changing the radius $r$ of the cylindrical resonator changes the scattering cross section spectrum. We can see that as $r$ increases the whole spectrum redshifts, in an approximately linear fashion. At the same time, the resonant peaks get broader, and due to this the scattering efficiency at the anapole increases, blurring it into the spectrum.

Changing the height $h$ instead of the radius has a more complex effect, as displayed on \autoref{fig:anapole_disp}.b. It shifts different modes in different ways, leading to weakly coupled modes crossing each other. This leads to, for example, a resonant mode crossing the anapole at $h=0.75$\textmu m and $\lambda=6$\textmu m, causing a dip in the scattering efficiency compared to the mode, but increasing it significantly compared to the anapole, making it vanish. As the mode redshifts at a higher rate compared to the anapole for increasing $h$, the anapole is restored at $h=1$\textmu m and $\lambda=6.5$\textmu m. At the same time, the scattering intensity increases at $h=0.85$\textmu m and $\lambda=7$\textmu m due to two modes coalescing and interfering constructively at a diabolical point~\cite{zhang_non-hermitian_2025}. As opposed to this, based on the trajectory of the modes in the region $6$\textmu m $<\lambda<8.5$\textmu m and $1.5$\textmu m $<h<2$\textmu m, the sudden increase in the scattering intensity at $h=2$\textmu m and $\lambda=8$\textmu m may potentially be due to superscattering effects~\cite{canos2023superscattering}, and can indicate the presence of an exceptional point~\cite{zhang_non-hermitian_2025}.

\newpage
\section{Light-matter interaction in a metamaterial disk}
\subsection{Maxwell-Bloch equations for a four-level system}

We use the well know Maxwell-Block equations, which provide a semi-classical description for the four-level system~\cite{valcarcel_semiclassical_2006,wuestner_gain_2011}. One can write the following system of coupled differential equations for the population density of the dipole emitter as
\begin{align}
\pdv{N_0}{t}&=\frac{N_1}{\tau_{10}}+\frac{N_3}{\tau_{30}}-\frac{1}{\hbar\omega_a} \vb E \vdot \left( \pdv{\vb P_a}{t} +\frac{\Gamma_a}{2} \vb P_a \right) \,, \\
\pdv{N_1}{t}&=-\frac{N_1}{\tau_{10}}+\frac{N_2}{\tau_{21}}-\frac{1}{\hbar\omega_e} \vb E \vdot  \left( \pdv{\vb P_e}{t} +\frac{\Gamma_e}{2} \vb P_e \right) \,, \\
\pdv{N_2}{t}&=\frac{N_3}{\tau_{32}}-\frac{N_2}{\tau_{21}}+\frac{1}{\hbar\omega_e} \vb E \vdot  \left( \pdv{\vb P_e}{t} +\frac{\Gamma_e}{2} \vb P_e \right) \,,\\
\pdv{N_3}{t}&=-\frac{N_3}{\tau_{32}}-\frac{N_3}{\tau_{30}}+\frac{1}{\hbar\omega_a} \vb E \vdot \left( \pdv{\vb P_a}{t} +\frac{\Gamma_a}{2} \vb P_a \right) \,,
\end{align}
and the polarization of the material
\begin{align}
\pdv[2]{\vb P_a}{t}+\Gamma_a \pdv{\vb P_a}{t} + \left(\frac{\Gamma_a^2}{4}+\omega_a^2 \right) \vb P_a = -\frac{\omega_a}{\hbar}\mu^2 (N_3-N_0)  \vb E \,, \\
\pdv[2]{\vb P_e}{t}+\Gamma_e \pdv{\vb P_e}{t} + \left(\frac{\Gamma_e^2}{4}+\omega_e^2 \right) \vb P_e = -\frac{\omega_e}{\hbar}\mu^2 (N_2-N_1)  \vb E \,,
\end{align}
where $N_j=\rho n_j$ denotes the absolute population density of level $j$, with $\rho$ being the number density of molecules and $0<n_j<1$ is the normalised population density, $\tau_{jl}$ is the non-radiative decay rate between levels $j$ and $l$, $\omega_a$ and $\omega_e$ are the absorption and emission transition frequencies, respectively, $\vb P_a$ and $\vb P_e$ is polarization corresponding to the absorption and emission transition, and $\Gamma_a$ and $\Gamma_e$ is the linewidth of the absorption and emission linewidth, and $\vb E$ is the electric field component parallel to the dipole, coupling the above system of equations to the wave equation, and finally $\mu$ is the transition dipole moment which we chosen to be the same for the emission and absorption for simplicity. We also define the coupling coefficient $\sigma=\mu^2 \rho/\hbar$.

\newpage
\subsection{Scattering cross section of a single layer aggregate}
\begin{figure}[h]
    \centering
    \includegraphics[width=0.5\linewidth]{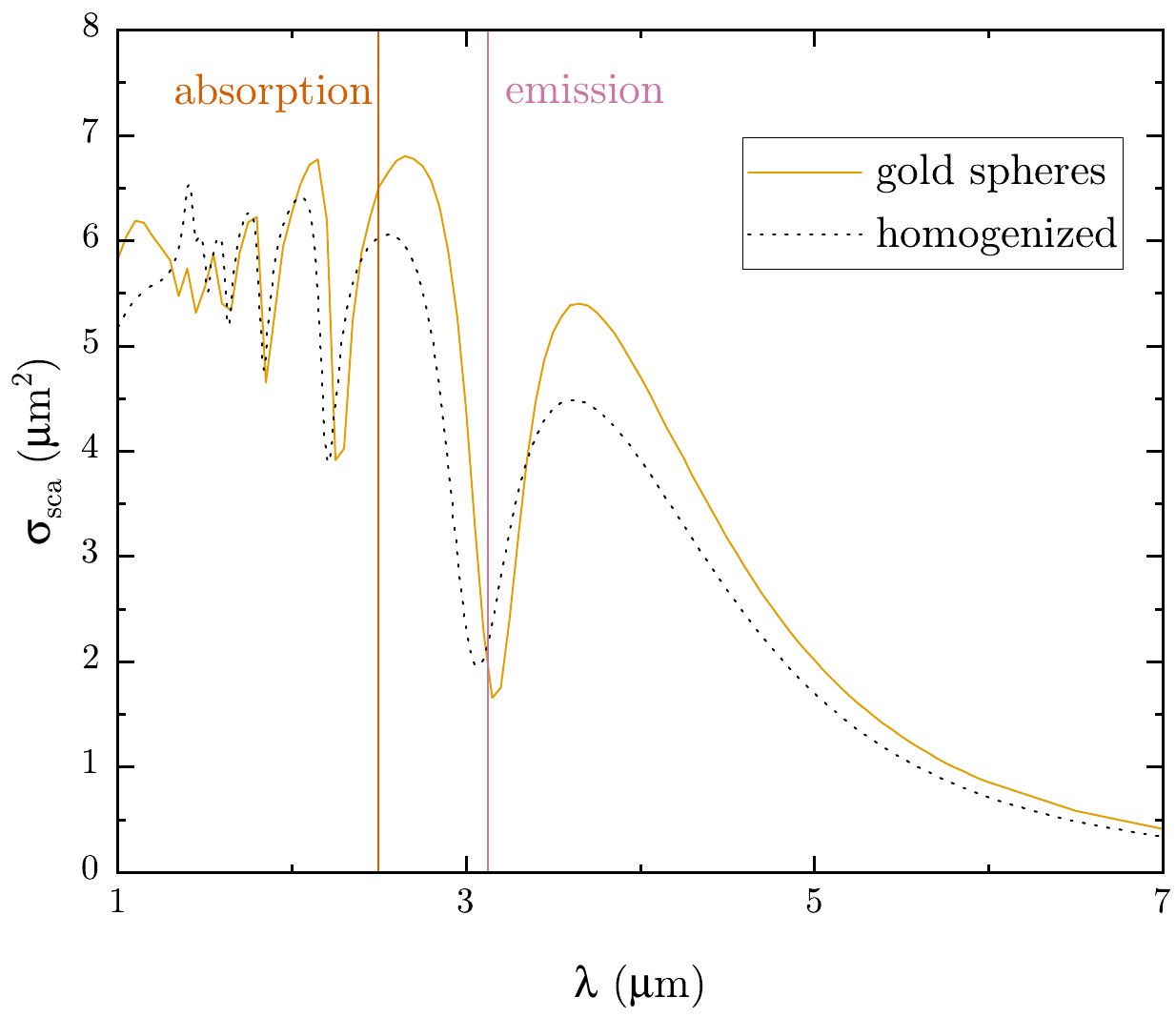}
    \caption{Scattering cross section spectra of a metamaterial disk of radius $R=1$\textmu m, and a height of $100$nm. Vertical lines indicate the transition frequencies in the Maxwell-Bloch equations.}
    \label{fig:scattering_1ML}
\end{figure}

To reduce computational complexity for the time domain calculation the thickness of the metamaterial disk resonator is reduced to a single layer of NP aggregates. Accordingly, the spectra is blueshifted, see \autoref{fig:anapole_disp} for dispersion and  \autoref{fig:scattering_1ML} for the detailed spectra of a single NP aggregate layer disk resonator.

\newpage
\subsection{Emission modes}
\begin{figure}[h]
    \centering
    \includegraphics[width=0.7\linewidth]{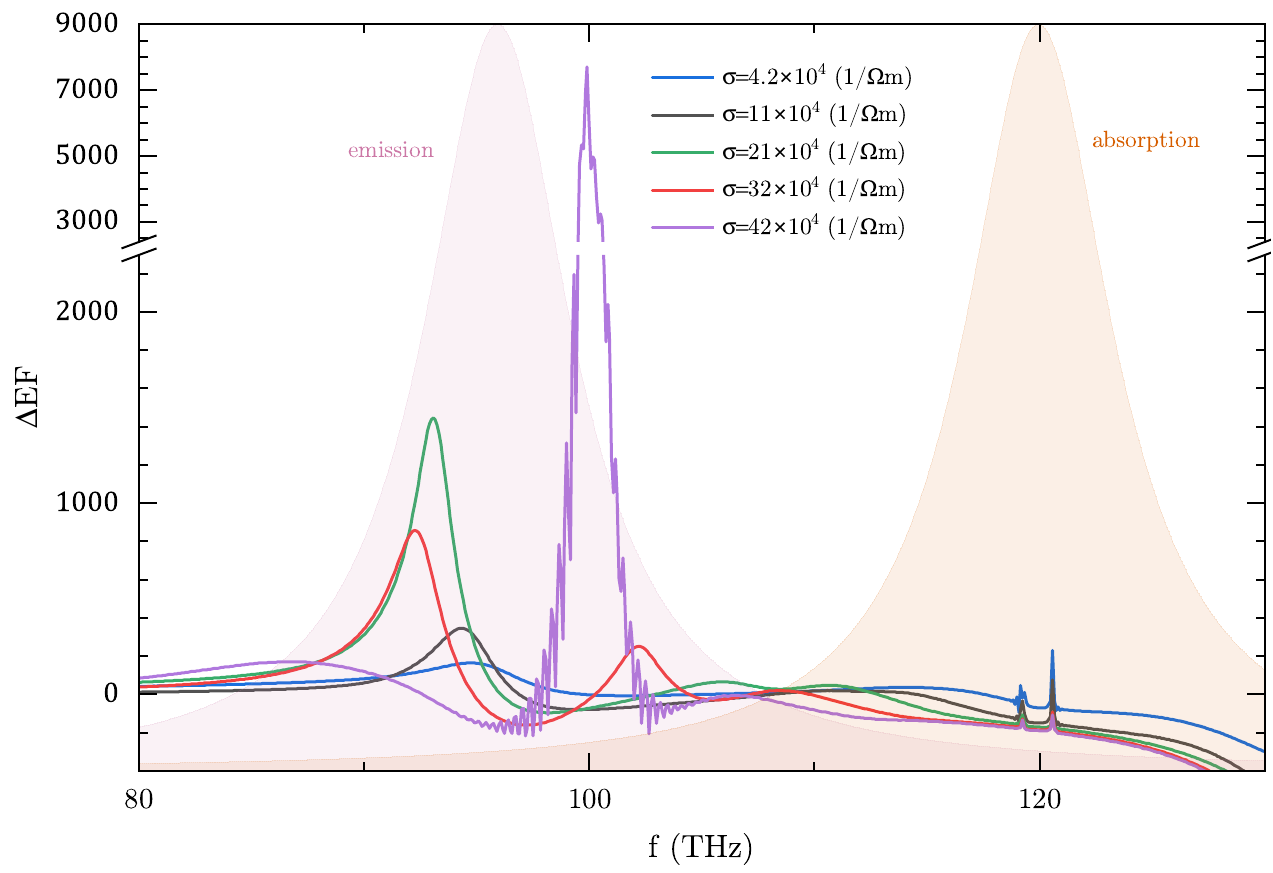}
    \caption{Difference of enhancement factor in a gap near the center of the disk resonator. Note the cut of the $y$ axis. The sharp peak around $f=120$THz, and the oscillations around the peak at $f=100$THz are due to numerical noise in the Fourier-transform of the data arising from termination of time domain simulation.}
    \label{fig:mode_switch}
\end{figure}

The anapole arises as an interference between two eigenmodes and it in the vicinity of  an exceptional point in parameter space~\cite{zhang_non-hermitian_2025}. Due to the vicinity of the exceptional point, as we pump the system harder we find emission mode switching~\cite{fischer_controlling_2024} from $f=92$THz at $\sigma=5.0$(C/m\textsuperscript{2}) to $f=102$THz 
at $\sigma=6.7$(C/m\textsuperscript{2}), see \autoref{fig:mode_switch}.

\begin{figure}[h]
    \centering
    \includegraphics[width=0.9\linewidth]{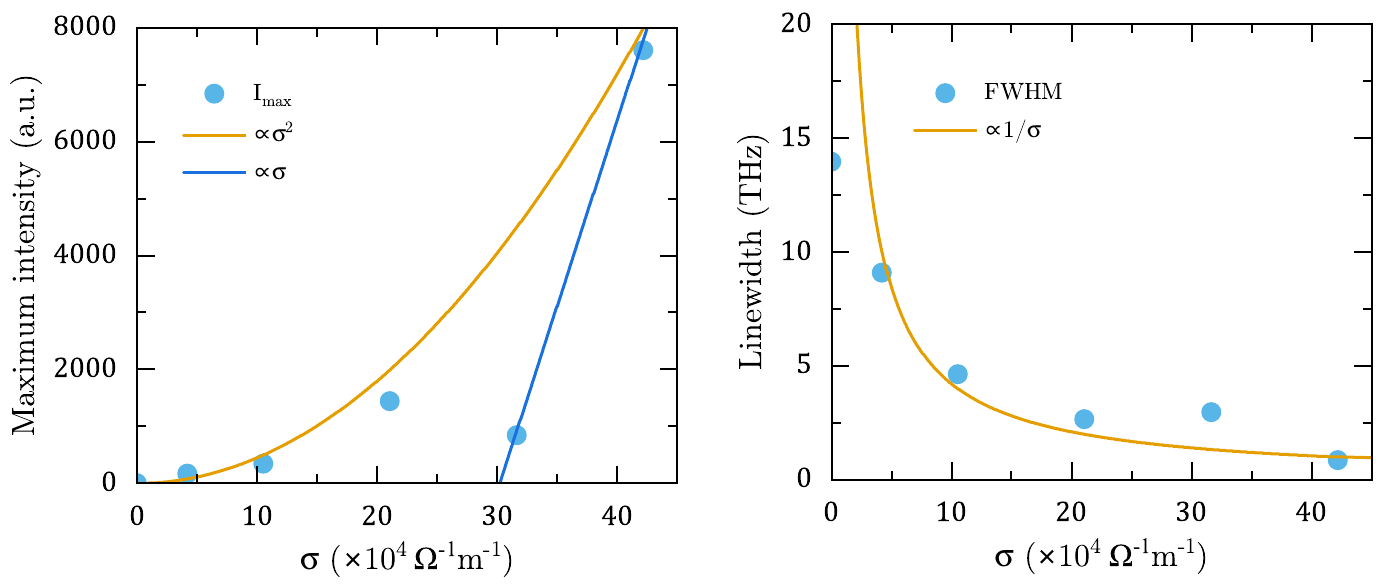}
    \caption{Intensity and linewidth of emission. Solid lines are reference curves.}
    \label{fig:linewidth_intensity}
\end{figure}

On \autoref{fig:linewidth_intensity} we show the intensity and linewidth change as we increase the pumping of the system. We observe an increase of intensity and decrease of linewidth with the pumping, indicating the onset of lasing, with the only deviation occurring due to the mode switching around 
$\sigma=32\times10^4$(1/$\Omega$m).

\printbibliography[heading=subbibliography,title={References}]
\end{refsection}

\end{document}